\documentclass[useAMS,usenatbib]{mn2e}
\usepackage{graphicx}
\usepackage[colorlinks=true,citecolor=blue,linkcolor=blue]{hyperref}
\usepackage{amssymb,natbib}

\usepackage{fancyhdr}
\usepackage{longtable,lscape,amsmath}  
\usepackage{rotating}
\usepackage{threeparttable}
\usepackage{xcolor}
\usepackage{natbib,twoopt}
\usepackage[T1]{fontenc}
\usepackage{aecompl}

\newlength{\onecolplotw}
\newlength{\twocolplotw}
\def\kmss    {km\,s$^{-1}$}
\def\kms     {km\,s$^{-1}$ }
\def\kmspc     {km\,s$^{-1}$pc$^{-1}$}
\def\Jyb    {Jy\,beam$^{-1}$}

\begin{document}
\bibliographystyle{mn2e}

 \title[The OHMM in Arp\,220: the rest of the story]{The OH Megamaser Emission in Arp\,220: the rest of the story}

\author[Baan, Aditya, An, and Kl\"ockner]
{Willem A.~Baan$^{1,2,3}$\thanks{E-mail: \href{mailto:baan@astron.nl}{baan@astron.nl}},  
J.N.H.S. Aditya$^{3,4,5}$, 
Tao~An$^{3}$, 
and 
Hans-Rainer Kl\"ockner$^{6}$
\vspace{5mm}
\\
$^1$Netherlands Institute for Radio Astronomy (ASTRON), 7991 PD Dwingeloo, the Netherlands \\
$^2$Xinjiang Astronomical Observatory, Chinese Academy of Sciences, 150 Science 1-Street, 830011 Urumqi, China\\
$^3$Shanghai Astronomical Observatory, Chinese Academy of Science, 80 Nandan Road, Shanghai 200030, China \\
$^4$Sydney Institute of Astronomy, School of Physics A28, University of Sydney, NSW 2006, Australia \\
$^5$ARC Centre of Excellence for All Sky Astrophysics in 3 Dimensions (ASTRO 3D), Mount Stromlo Rd, Australia Capital  Territory, 2611 AU, Australia \\
$^6$Max-Planck-Institut f\"ur Radioastronomie, Auf dem H\"ugel 69, 53121 Bonn, Germany 
}
\newcommand{\orcidauthorA}{0000-0003-3389-6838} 
\newcommand{\orcidauthorB}{0000-0002-0268-0375} 

   \date{Draft: \today}
   
\date{Accepted xxx. Received xxx; in original form xxx}
\pagerange{\pageref{firstpage}--\pageref{lastpage}} \pubyear{2023}

   \maketitle   
   
\begin{abstract} 
The OH Megamaser emission in the merging galaxy Arp\,220 has been re-observed with the Multi-Element Radio Linked Interferometer Network (MERLIN) and the European VLBI Network (EVN). 
Imaging results of the OH line emission at the two nuclei are found to be consistent with earlier observations and confirm additional extended emission structures surrounding the nuclei.
Detailed information about the distributed emission components around the two nuclei has been obtained using a concatenated MERLIN and EVN database with intermediate (40 mas) spatial resolution. 
Continuum imaging shows a relatively compact West nucleus and a more extended East nucleus in addition to an extended continuum ridge stretching below and beyond the two nuclei. 
Spectral line imaging show extended emission regions at both nuclei together with compact components and additional weaker components north and south of the West nucleus. 
Spectral line analysis indicates that the dominant OH line emission originates in foreground molecular material that is part of a large-scale molecular structure that engulfs the whole nuclear region. 
Compact OH components are representative of star formation regions within the two nearly edge-on nuclei and define the systemic velocities of East and West as 5425 \kms and 5360 \kmss.
 The foreground material at East and West has a 100 \kms lower velocity at  5314 and 5254 \kmss.
These emission results confirm a maser amplification scenario where the background continuum and the line emission of the star formation regions are amplified by foreground masering material that is excited by the FIR radiation field originating in the two nuclear regions.
\end{abstract}

\begin{keywords} masers  $-$  ISM: molecules  $-$  
             radio lines: ISM   $-$ galaxies:ISM $-$ galaxies: nuclei $-$ $-$ radio lines: galaxies
\end{keywords}

%

\section{Introduction}

The interacting system IC\,4553/4, also known as Arp\,220, is the host galaxy system of the first known  hydroxyl (OH) MegaMaser (OHMM; \citealp[][]{BaanWH1982}). The OH emission in Arp\,220 was discovered at the Arecibo Observatory during a search for OH absorption in sources with strong HI absorption \citep{Mirabel1982}.
The extraordinary properties of Arp\,220 were later confirmed by the far-infrared (FIR) prominence in the Infrared Astronomical Satellite (IRAS) data \citep{Soifer1984,SandersEA1988}, and the source became the prototype of the Ultra-Luminous InfraRed Galaxy (ULIRG) population \citep{SandersM1996}.
Subsequent searches for OHMM emission have resulted in about 120 systems among the ULIRG population \citep{Baan1989, DarlingG2002,Kloeckner2004,ZhangEA2014}. 

Arp\,220 is a merger system with two radio nuclei \citep{BaanH1995} embedded in a chaotic optical structure \citep{LockhartEA2015}.
The ongoing merger has triggered a powerful burst of star formation at each of the nuclei \citep{1997ApJ...484..702S}, which results in the FIR prominence and multiple radio supernova remnants (SNRs) at each of the nuclei \citep{SmithEA1998,LonsdaleEA2006,VareniusEA2019}, as well as starburst (SB)-related hard X-ray emission \citep{ClementsEA2002,IwasawaEA2005}. 
Arp\,220 is a (nearby) template for high redshift active galaxies with short-lived bursts of assembly and nuclear activity that appear to be common at redshifts of $\sim$2.5 and define the characteristics of massive galaxies in the nearby Universe. 

The radio positions of the two nuclei in Arp\,220 are separated by 0.97\arcsec\  (371 pc) on the plane of the sky \citep{BaanH1995,DownesS1998,SakamotoEA1999}, while the optical images show a dust-enshrouded structure \citep{WilsonEA2006,LockhartEA2015}. 
A radio continuum bridge between and below the two nuclei is the only evidence for the interactive nature of the system. 
However, because of the puzzling nature of the system, the systemic velocities of the two underlying nuclei remain to be determined.
A reinterpretation of the early Very Large Array (VLA) A-array observations of the 1667 and 1665 MHz OH emission in Arp\,220 would suggest that the features of the East and West nuclei are merged within the observed spectral signatures and that the velocities of West and East are close to 5350 \kms and somewhat higher than 5390 \kmss, respectively \citep{BaanH1984}. 
Early Multi-Element Radio-Linked Interferometer Network (MERLIN) observations confirm that the 1667 MHz emission at the velocity of the Western nucleus appears close to the Eastern nucleus and that the velocity fields of the two nuclei may be mixed \citep{RovilosEA2003}.
The distribution of the CO emission also indicates that the velocities at the West and East nuclei are approximately at 5370 and 5400 \kms \citep{WheelerEA2020,SakamotoEA2008,RangwalaEA2015}.
A study of the dynamics of Arp\,220 based on early detection of formaldehyde and the corresponding OH emission employed similar velocities of 5346  and 5418 \kms at the West and East nuclei for understanding the nuclear antics of the system  \citep{BaanH1995}.
The formaldehyde emission in Arp\,220 is found to extend across the central molecular zone of each of the nuclei and covers the systemic velocities of both nuclei  \citep{BaanEA2017}. 
Arp\,220 also exhibits an OH outflow feature that extends to $\sim$1000 \kms below the OH 1667 MHz feature \citep{BaanHH1989}.

The observed OH MM emission has been interpreted with an amplification scenario where foreground excited and masering material amplifies the background radio continuum \citep{Baan1989}. 
The OH emission would thus be superimposed on the radio structure of the source and the FIR emission regions generated by dust emission resulting from ongoing star formation, which has been suggested to serve as a pumping agent for the OH molecules in the foreground. 
Both compact high-brightness and extended low-brightness maser components could ensue in this manner.

The re-observations of the Arp\,220 system of the complete 1667 and 1665 MHz emission spectrum presented in this paper have been taken with MERLIN and with the European VLBI Network (EVN) . 
Previous interferometric MERLIN and EVN observations only covered the prominent 1667 MHz OH emission originating at both nuclei. 
The lower resolution images from the MERLIN observations provide an integrated view of the OH emission in Arp\,220 without fully detailing the structural components of Arp\,220.
The global Very Long Baseline Interferometry observations with the EVN provide a highly resolved view of the nuclear regions and only found compact emission components.
In order to identify and image the two nuclei in more detail, the two data sets will be concatenated, which will give a data base with intermediate resolution and allows mapping the spatial structure of both the 1667 and 1665 MHz OH emission regions. 
And this new database reveals some of the hidden secrets of the Arp\,220 system.

\begin{figure}
\begin{center}
\includegraphics[width=0.9\columnwidth,angle=-90]{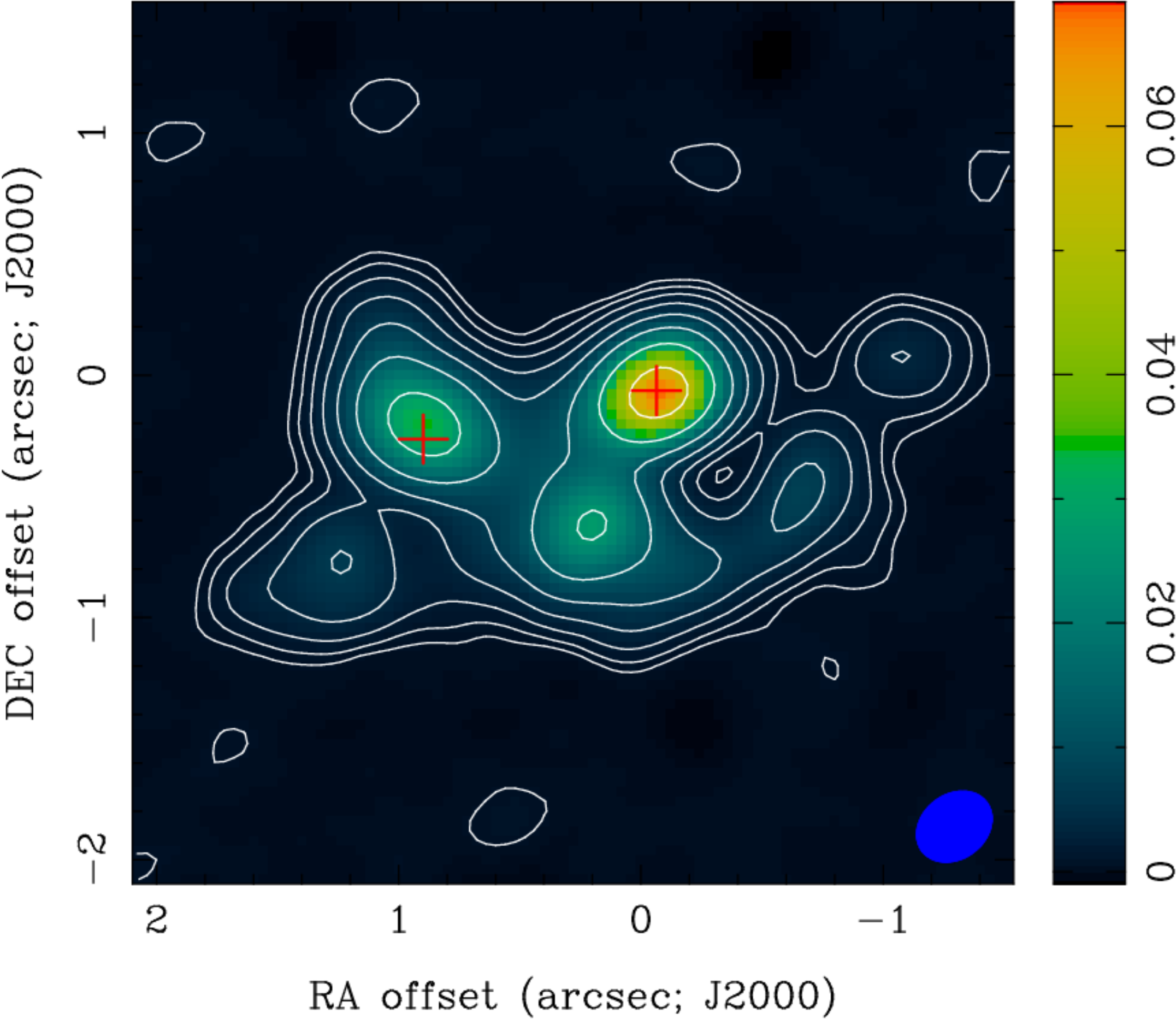}
\caption{The 1.6-GHz continuum emission of Arp\,220 made from the MERLIN data. 
The restoring beam is 0.28" $\times$ 0.26", PA=$-$63.6$^o$. 
The peak intensity is 53.5 mJy beam$^{-1}$ and the rms noise in the off-source region is 0.3 mJy beam$^{-1}$. The contours are at 0.80 mJy beam$^{-1}$ $\times$(1, 2, 4, 8, 16, 32, 64). 
The two crosses representing the dust emission peak positions at 230 GHz \citep{SakamotoEA1999} are in excellent agreement with the present continuum peaks.}
\label{fig1}
\end{center}
\end{figure}

\begin{figure*}
\begin{center}
\includegraphics[width=1.2\columnwidth,angle=0]{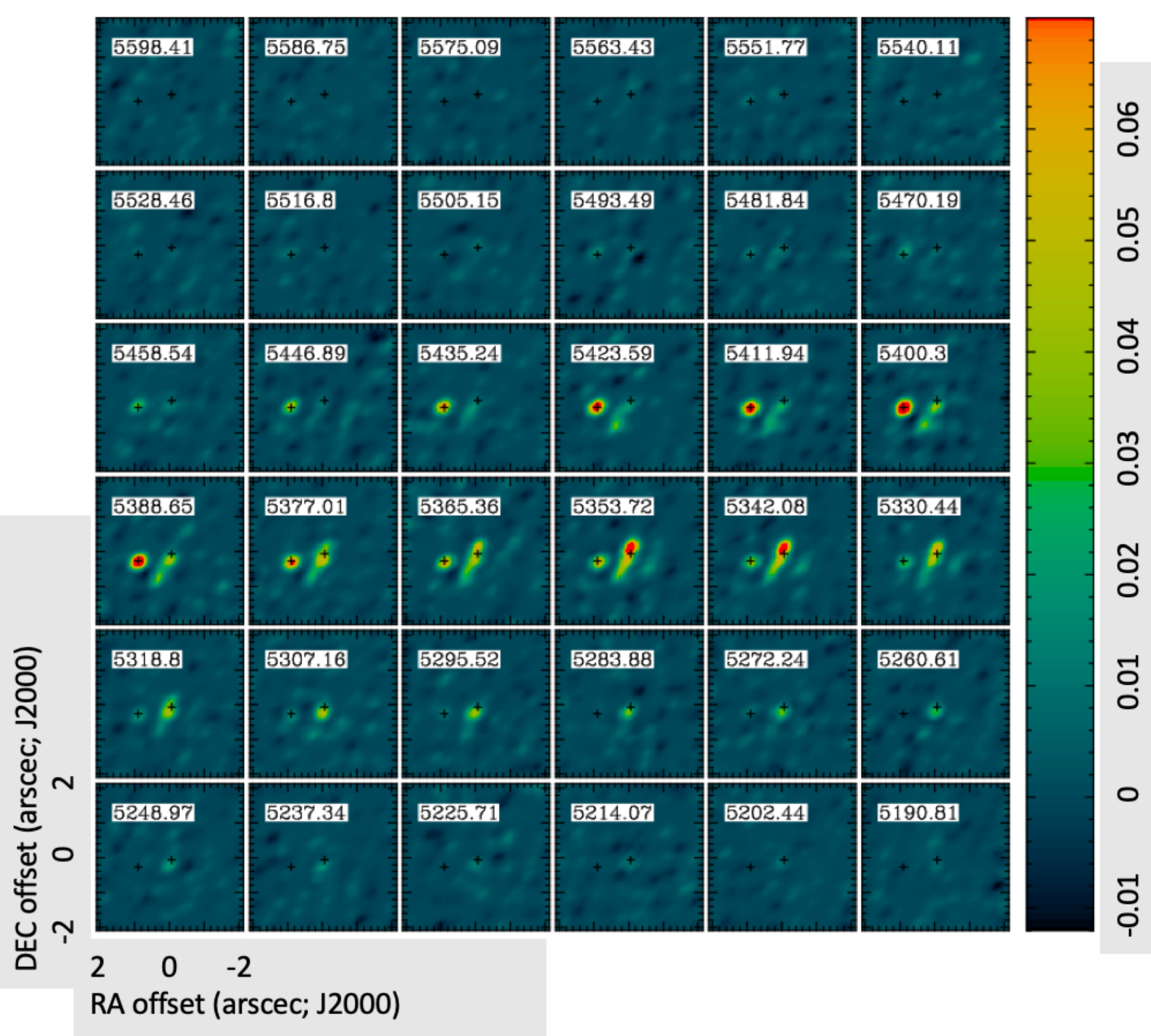}
\caption{Channel maps of the OH 1667 line emission from the MERLIN data. 
The restoring beam is 0.23" $\times$ 0.19" at a PA=32$^o$. 
The velocity scale is based on the rest frequency of the 1667 MHz transition.
The rms noise level is 2.0 mJy beam$^{-1}$. 
The peak intensity is 157 mJy beam$^{-1}$ in the 5363.5 \kms channel. 
For clarity, only the contour of 6 mJy beam$^{-1}$ is plotted in the image.}
\label{fig2}
\end{center}
\end{figure*}

\begin{figure*}
\begin{center}
\includegraphics[width=2.1\columnwidth,angle=0]{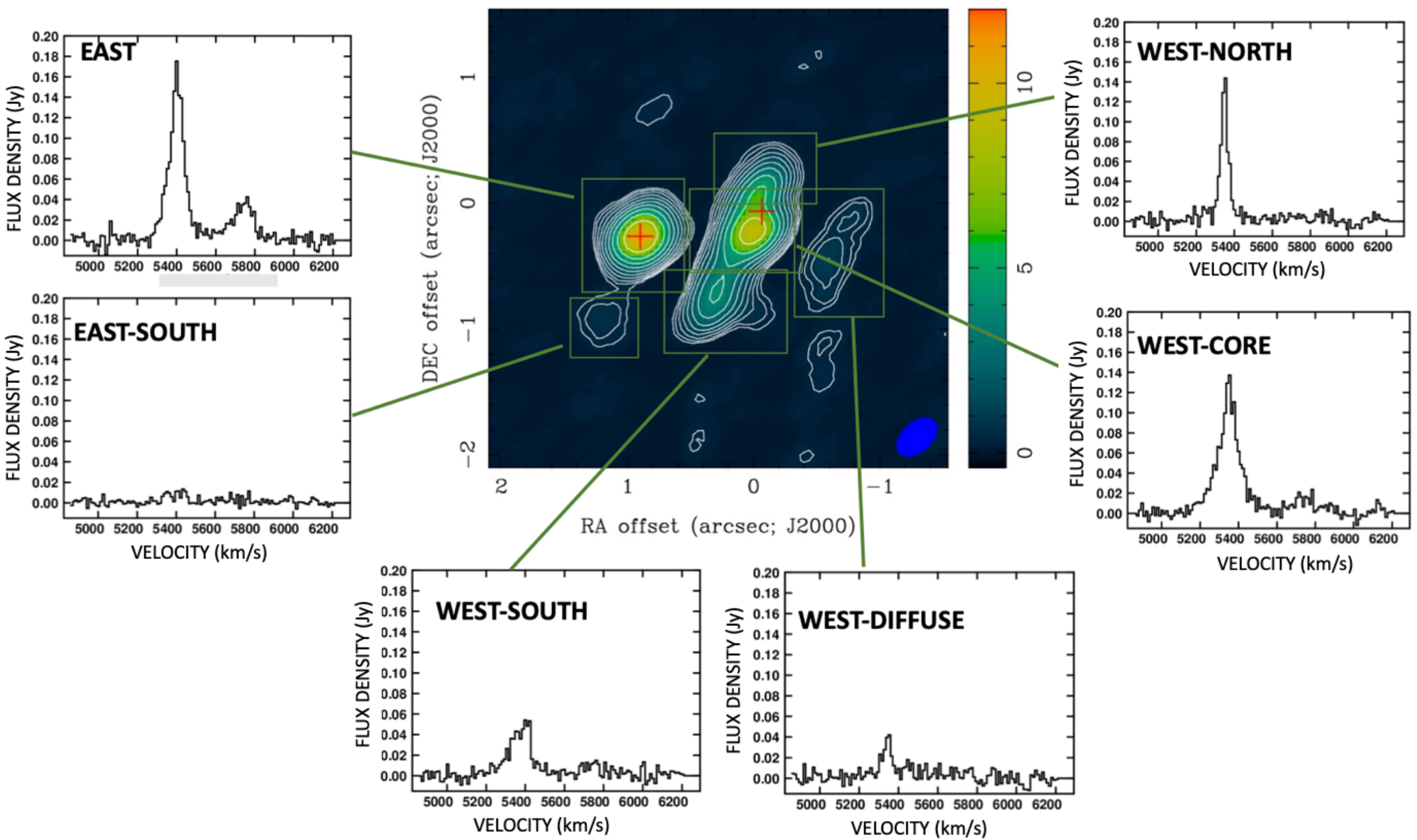}
\caption{The velocity-integrated intensity (moment 0) map of the OH 1667 line emission made by integrating the line emission with the signal-to-noise ratio higher than 3 in each velocity channel.
The beam for these data is 0.23" $\times$ 0.19". 
Two crosses mark the position of the two continuum peaks. 
The peak intensity is 15.17 \Jyb \kms and the colour scale is in  \Jyb \kmss.
The contours are 0.5\%, 2\%, 5\%, 10\%, 20\%, 30\%, ...90\% of the peak. 
The integrated spectra of the 1667 and 1665 MHz line emission for four selected regions are shown as side boxes using the velocity registration of the 1667 MHz line.
The y-axes for the spectral plots are in units of Jansky.
The heliocentric velocity scale runs from 4910 to 6300 \kms and increases from left to right.
The integrated flux densities of the components are: 
East component: 20.00 +/- 4.81 Jy beam$^{-1}$,
West-South component: 14.62 +/- 4.61 Jy beam$^{-1}$, and 
West-North component: 7.45 +/- 4.66 Jy beam$^{-1}$. }
\label{fig3}
\end{center}
\end{figure*}

\section{Observations and Data Reduction}\label{sec2}

Throughout this paper a Hubble constant $H_0$ = 70 \kms Mpc$^{-1}$ has been assumed, which indicates that for the Arp\,220 system, the angular-spatial size conversion results in 1 arcsecond corresponding to 382 pc.
All results are presented using the optical helio-centric definition of velocity. 
For Arp\,220, velocities are about 96 \kms lower using the radio definition.

\subsection{MERLIN Observations and Data Reduction}\label{sec21}

The MERLIN observations of Arp\,220 were carried out on 2003 June 24th and 25th with all seven MERLIN antennas, including the 76-metre Lovell telescope in left-hand circular polarisation mode. 
The data of the observing project MN/03B/22 were recorded in 128 channels covering a total bandwidth of 8 MHz, each with a channel width of 62.5 kHz, giving a velocity resolution of 11.66 \kms at 18 cm wavelength. 
The sources 3C\,84, 3C\,286, and J\,1516+1932 were used as bandpass, flux density scale, and phase referencing calibrators, respectively. The whole observation of two sessions lasted 18 hours, two third of which was spent on Arp\,220.

The preliminary amplitude calibration, bandpass and antenna-based phase calibration were made in the Astronomical Image Processing System (AIPS) using the calibrator sources. 
The gain solutions derived from the calibrators were applied to the Arp\,220 data by interpolation. 
Then the Arp\,220 data were exported out of the multi-source dataset and were imported into MIRIAD for self-calibration and imaging.

The data were first averaged in channels to produce a single-channel dataset, a so-called pseudo-continuum dataset. 
An image of Arp\,220 was created using the pseudo-continuum data. 
After flagging some discrete bad data points induced by radio frequency interference (RFI) and other observational problems, this pseudo-continuum data was used for a few iterations of phase-only self-calibration with the time intervals starting from 5 minutes to 0.5 minutes until the dynamic range of the CLEANed image did not improve any more.  
A final image was produced with natural weighting (see Figure \ref{fig1}). 

The antenna gains as a function of time determined in the self-cal procedure were applied to the line data. 
The Miriad task UVLIN was used to subtract the continuum emission from the visibility data by fitting a polynomial to the real and imaginary parts of the selected line-free channels across the line cube. 
The continuum and line data are employed separately to make the resulting maps.

\subsection{EVN Observations and Data Reduction}
\label{sec22}

Arp\,220 was observed with the EVN from 2003 June 24th UT18:00 to June 25th UT03:30 with observing program EB022C.
Twelve telescopes participated in the observations: Jodrell Bank, Effelsberg, Cambridge, Noto, Torun,  Shanghai, Westerbork, Onsala, Medicina, Urumqi, Hartebeesthoek and Robledo. 
The observations were made in dual circular polarisation mode and the data were recorded in 256 channels. 
The total bandwidth is 8 MHz; therefore, each channel has a width of 31.25 kHz (corresponding to a velocity  resolution of 5.83 \kms at 18 cm wavelength). 
OQ\,208 was used as the bandpass calibrator and J\,1613+3412 and J\,1516+1932 were used as phase referencing calibrators. The whole observation lasted for 9.5 hours, during which Arp\,220 was observed for 6 hours. 

The calibration of the multiple-source dataset was made in AIPS. 
The data were first sorted in time-baseline sequence. 
The amplitude calibration was done using the system temperatures of the observations and the gain curves provided by each station. 
The phase errors induced by the ionospheric effects are corrected using the AIPS task TECOR. 
Fringe fitting was carried out with the compact and strong calibrators, and the derived solutions were then applied  to the whole dataset to calibrate the delays, delay rates, and phases. 
Complex bandpass solutions were determined using the OQ\,208 data. 
RFI was identified in the total power spectra and the affected channels were flagged. 
The phase, gain, and bandpass solutions derived from the calibrators were applied to Arp\,220 data by interpolation. 
The source-rest-frame frequency was set to the line data and AIPS task CVEL was used to determine the Doppler shift  correction on each baseline.
The visibility amplitudes of the EVN calibrator data J\,1516+1932 and OQ\,208 were compared with those of the MERLIN data on common baselines, and they show consistency within 2 per cent. 
Considering the scattering of the EVN visibility amplitude and the variation of the telescope performance, we conservatively adopt an amplitude uncertainty of 5 per cent.

The calibrated data were exported from AIPS and imported into Miriad for further analysis. 
The line data were separated from the original data by using the task UVLIN, which fits the continuum emission with a linear function using line-free channel data and subtracts the fitted baseline from the line channels. 
Next, an iteration of self-calibration was applied to the line data on the line peak channels assuming a point source model. 
The line cubes were all mapped with natural weighting.

\subsection{Combining the MERLIN and EVN data}
\label{sec23}

A combination of MERLIN data with 280 x 260 mas with the EVN data set with resolution 10 x 10 mas would result in a data set with an intermediate resolution such that both the extended and the compact components in Arp\,220 are both identifiable in a single image.

In order to further optimise the available data, the residual phase errors on the short (and most sensitive) EVN baselines and the alignment of the EVN and MERLIN phase centres were corrected by using an EVN continuum calibration iteration employing a reference model formed by the CLEAN models derived from the MERLIN continuum data.

The images resulting from the combined MERLIN and EVN (ME) data have a restoring beam of $41 \times 38$ mas that will appropriately reveal the more extended emission regions at the nuclei as well as the very compact VLBI components.

\begin{figure}
\begin{center}
\includegraphics[width=0.7\columnwidth,angle=-90]{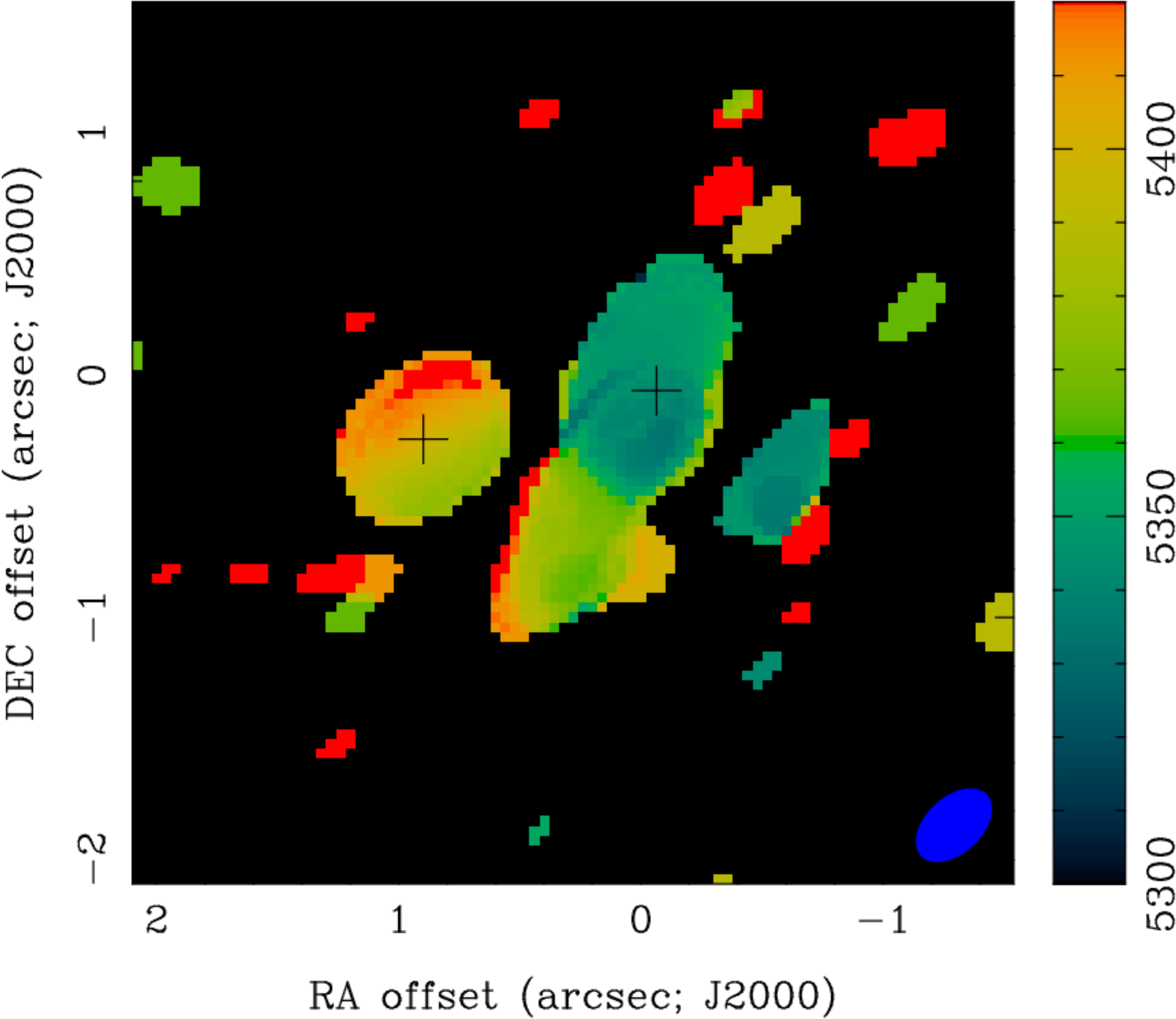}
\includegraphics[width=0.7\columnwidth,angle=-90]{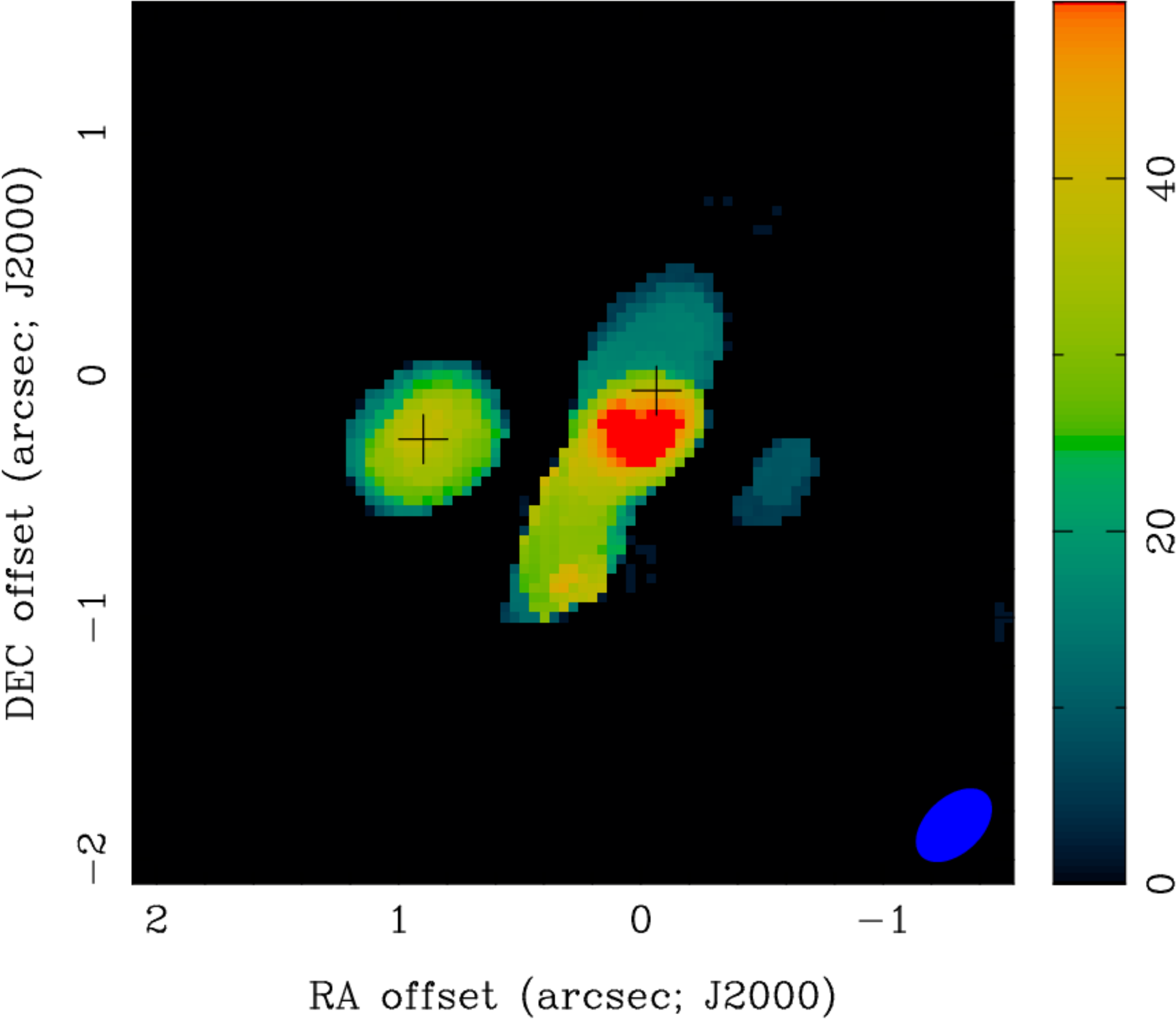}
\caption{The MERLIN maps or the 1667 MHz OH emission at both nuclei.  
(top) The velocity distribution (1Moment) and (bottom) the velocity width (2Moment) maps. 
The colour scale are in units of \kmss. }
\label{fig4}
\end{center}
\end{figure}

\begin{figure}
\begin{center}
\includegraphics[width=0.85\columnwidth,angle=-90]{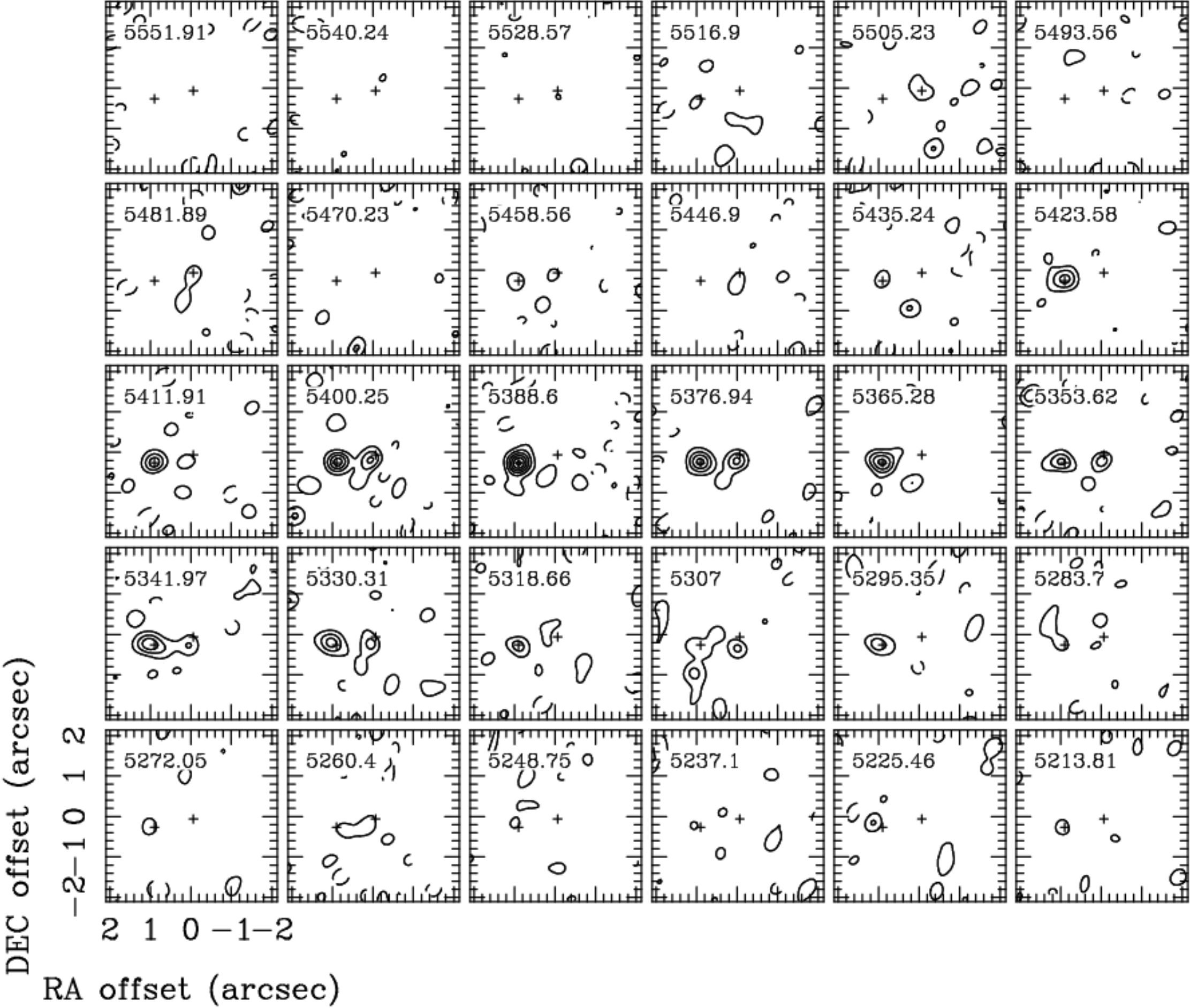}
\caption{Channel maps of the OH 1665 line emission from the MERLIN data. 
Beam = 0\farcs23 $\times$ 0\farcs19, PA = 32$\degr$. 
The velocity scale is based on the rest frequency of the 1665 MHz transition.
Contours: 6 mJy beam$^{-1}$ $\times$ (-1,1,2,3,4,5,6,7,8). 
The first contour represents 3 times the average off-source rms noise of 2 \Jyb. 
The peak intensity is 30.8 \Jyb in the 5376 \kms channel. 
Two crosses mark the position of two continuum peaks.
The restoring beam for these maps is 0.23" $\times$ 0.19".}
\label{fig5}
\end{center}
\end{figure}

\begin{figure}
\begin{center}
\includegraphics[width=0.7\columnwidth,angle=-90]{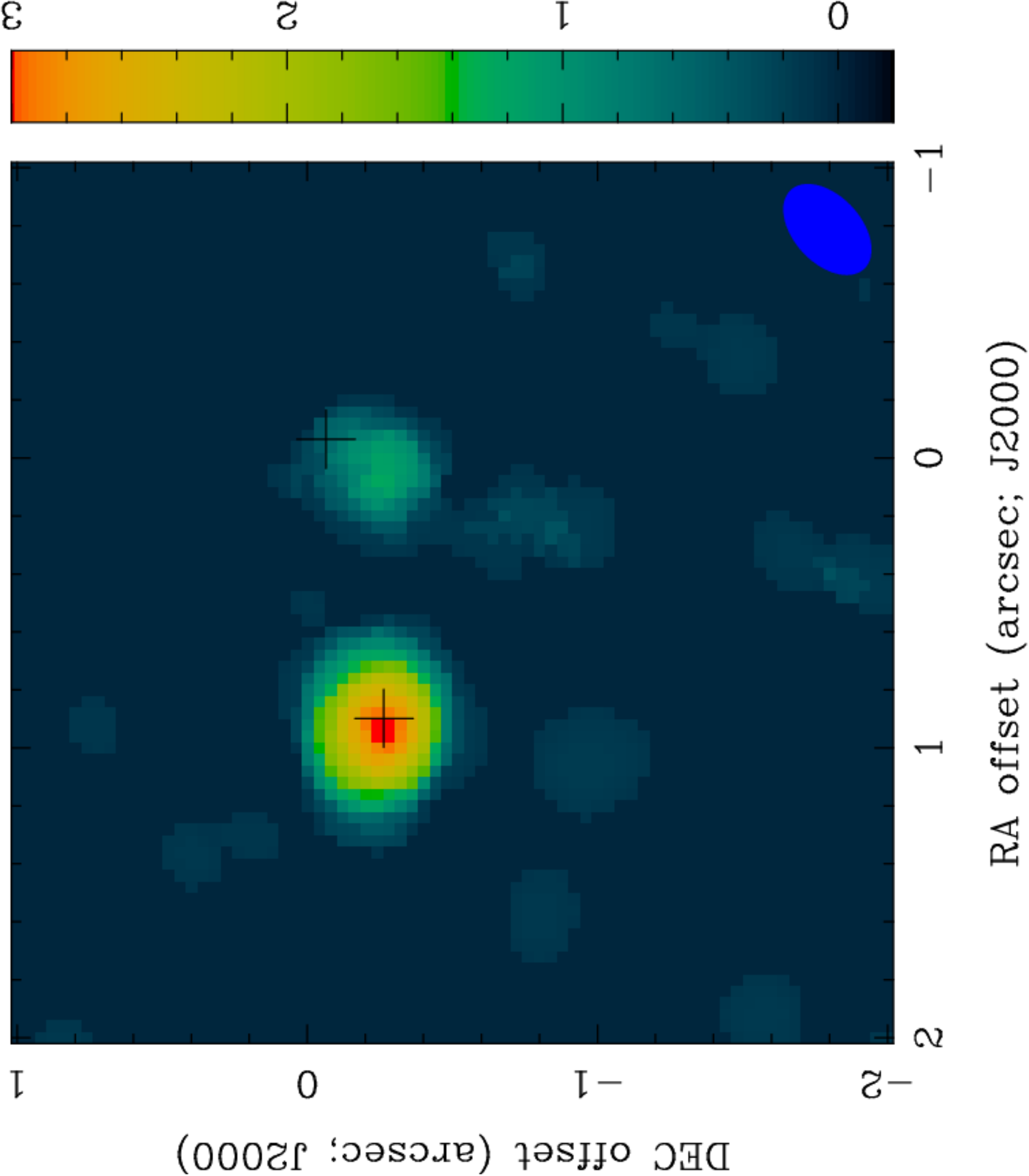}
\includegraphics[width=0.7\columnwidth,angle=-90]{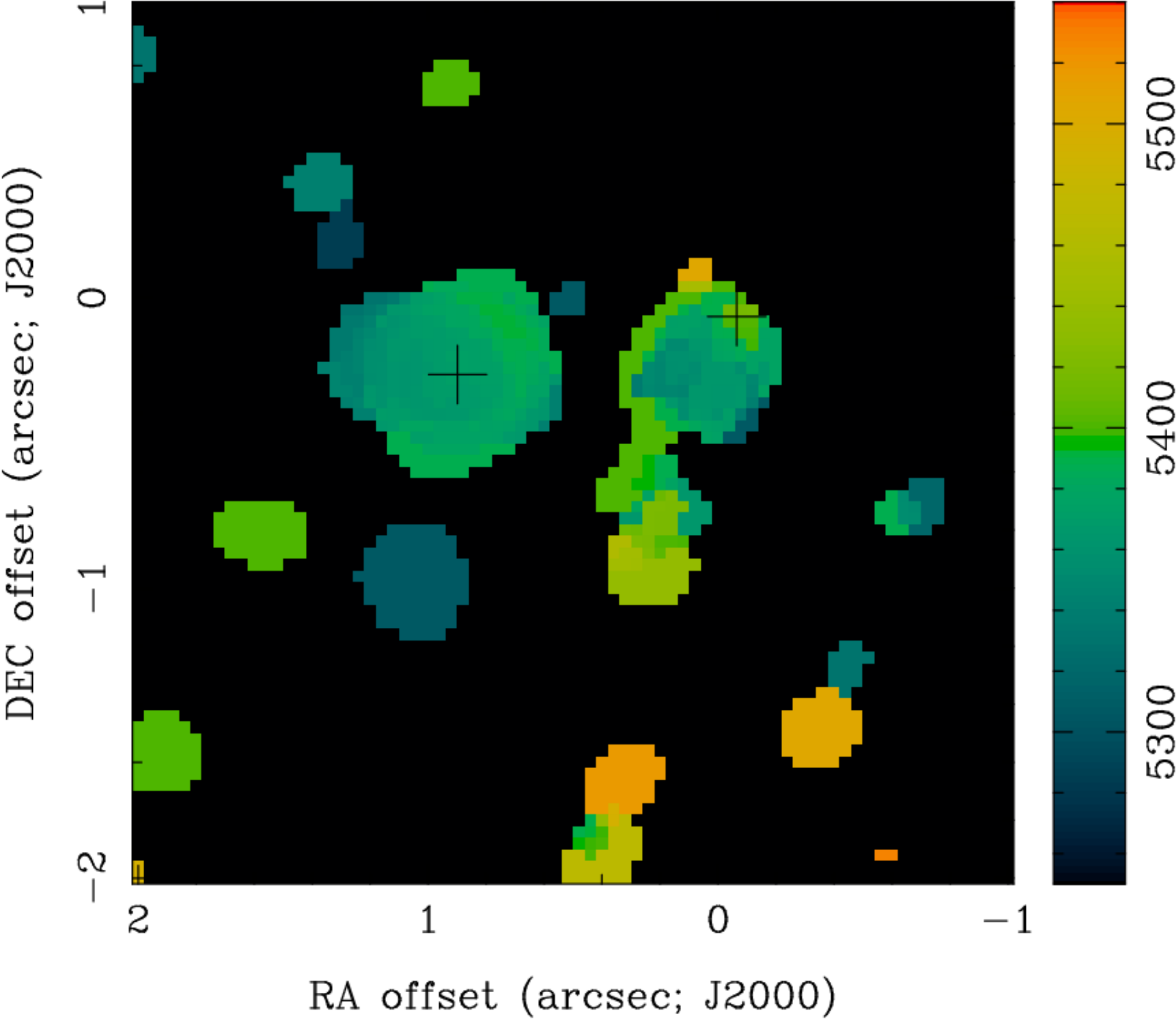}
\caption{The velocity-integrated 0Moment map (top) and 1Moment velocity map (bottom) of the OH 1665 line emission from the MERLIN data.}
\label{fig6}
\end{center}
\end{figure}

\section{MERLIN and EVN Imaging Results}\label{sec3}

Throughout this paper a Hubble constant $H_0$ = 70 \kms Mpc$^{-1}$ has been assumed for interpreting the imaging results.
For the Arp\,220 system with a systemic velocity of about 5400 \kmss, the angular-spatial size conversion results in 1 arcsecond corresponding to 375 pc.
All results are presented using the optical heliocentric definition of velocity. 
Other publications may use velocities using a radio definition, which for Arp\,220 are about 96 \kms lower.

Throughout the paper the continuum and spectral line images show two crosses representing the dust emission peak positions of the two nuclei at 230 GHz \citep{SakamotoEA1999}, which are in agreement with the continuum peaks from the current data sets.

\subsection{MERLIN Imaging of the Continuum Emission}\label{sec31}

The MERLIN continuum emission displays a double-component structure embedded within a larger envelope (Fig. \ref{fig1}).  
While the two prominent nuclear emission regions are found to be similar to those seen in the earlier less-sensitive MERLIN map \citep{RovilosEA2003}, the surrounding structures are much more detailed. 
The two main peaks, separated by about 0\farcs97 (365 pc), are in agreement with other published high-resolution images \citep[e.g.][]{SakamotoEA1999,RovilosEA2003,BaanEA2017}. 
The total flux density is estimated to be 274$\pm$15.2 mJy, in agreement with previous measurements at the same frequency \citep{RovilosEA2003}. 
The West nucleus is brighter and relatively more compact than the East nucleus.  
Gaussian fitting with a zero baseline gives an integrated flux density of 124.2$\pm$5.4 mJy and 99.4$\pm$3.4 mJy for the West and East components, respectively (Table \ref{tab1}).

However, the new MERLIN continuum map of Arp\,220 also shows prominent extensions to the south, the southeast, and particularly to the west.
The southern bridge below the two nuclei observed in previous data \citep{BaanH1995, RovilosEA2003} 
now appears as a continuous structure extending from southeast of Arp\,220E all the way to west of Arp\,220W. 
With a peak intensity of 13.5 mJy beam$^{-1}$ this accounts for 45 times the off-source noise. 
This (arm-like) continuum bridge as well as other extensions may represent trails of star formation regions and debris resulting from the galaxy merger.
This may be confirmed by the detection of a large kpc-scale  structure, interpreted as a star forming disk, observed at low radio frequencies (150 MHz) with the international Low-Frequency Array (LOFAR) telescope \citep{VareniusEA2016}.

\begin{table}
\begin{center}
\caption{Continuum Emission from the MERLIN data}
\label{tab1}
\begin{tabular}{lccc}
\hline
Label    &   $S_{peak}$  &    $S_{integrated}$  &  maj, min, PA \\
             & (mJy/b)       & (mJy)                   & \\
\hline
Arp\,220W & 75.1$\pm$3.4  & 124.2$\pm$5.4 & 0.24"  0.17"108.2$^o$ \\
Arp\,220E  & 38.0$\pm$1.3  & 99.4$\pm$3.4  & 0.43"  0.23"  48.7$^o$ \\
\hline
\end{tabular}
\end{center}
\end{table}

\begin{figure}
\begin{center}
\includegraphics[width=1.02\columnwidth,angle=0]{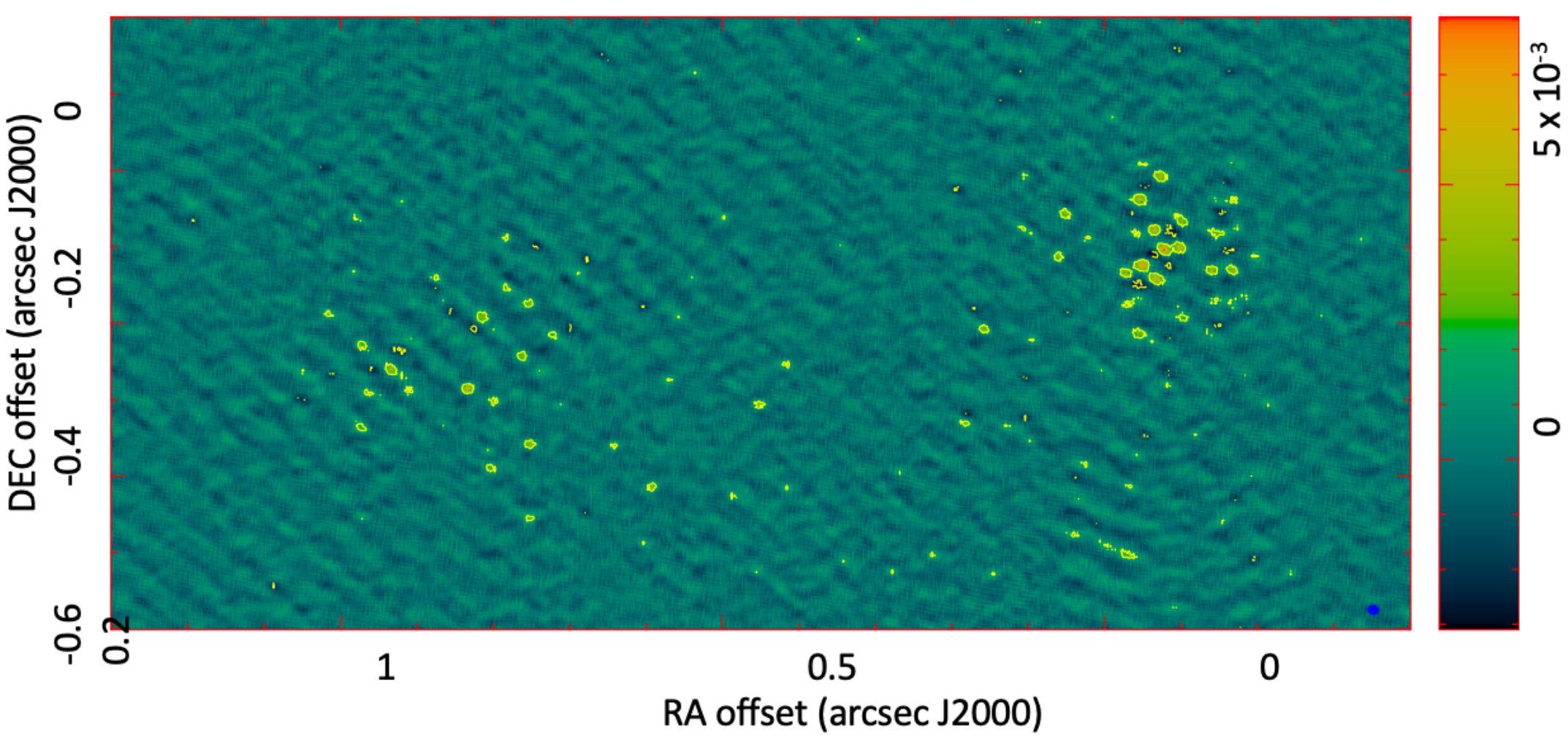}
\caption{The 1.6-GHz continuum emission of Arp\,220 made from the EVN data.
The restoring beam is 0\farcs013 $\times$ 0\farcs011, PA=74\fdg3. 
The peak intensity is 7.7 mJy beam$^{-1}$, and the rms noise in off-source regions is 0.5 mJy beam$^{-1}$.
The contour scale is in units of mJy beam$^{-1}$.}
\label{fig7}
\end{center}
\end{figure}

\subsection{The MERLIN OH 1667 Line Emission}
\label{sec32}

The MERLIN 1667 MHz OH emission line data cube of Arp\,220 covers a velocity range of 4900$-$6200 \kms with a velocity resolution of  11.7 \kmss. 
The 1667 MHz OH channel maps cover a range 5237 - 5447 \kms in Figure \ref{fig2} and show five distinct emission regions: the two nuclear regions, regions south and west of the West nucleus, and a region southeast of the East nucleus.
These structures  are clearly seen in the velocity-integrated 1667 MHz intensity map (grey scale and contours) together with the integrated 1665/1667 MHz spectra of six selected regions in the side panels  (Figure \ref{fig3}).
The current data provide more details of the emission structures at the two Arp\,220 nuclei than found in earlier MERLIN data, as those data did not incorporate the 1665 MHz line emission \citep{RovilosEA2003}.

Similar to the early data, the two OH components of Arp\,220W straddle the continuum peak, except that now the North component appears much less prominent than the South component, which has a peak intensity of 157 mJy beam$^{-1}$ in the 5363.6 \kms velocity channel.
The spectrum of the Arp\,220W-N region shows a narrow component at velocity 5353 \kms and is associated with a very compact masering region that also appears in the EVN data (see Section \ref{sec35}).
A third prominent component south of Arp\,220W is co-located with a  continuum component and appears more prominent than in the earlier data. 
This third component highlights the SE-NW orientation of both the continuum and line emission at Arp\,220W and suggests the orientation of the nuclear disk of the source in agreement with later findings (see Section \ref{sec4}).

The peak in Arp\,220E of 152 mJy beam$^{-1}$ is found in the 5410.2 \kms velocity channel at a position close to the Eastern continuum peak with the second peak in the 5375 \kms channel. 
The weaker components west and south of Arp\,220W and southeast of Arp\,220E seen in these high sensitivity data all lie within the confines of the continuum structure in Figure \ref{fig1}.

\begin{figure}
\begin{center}
\includegraphics[width=0.7\columnwidth,angle=0]{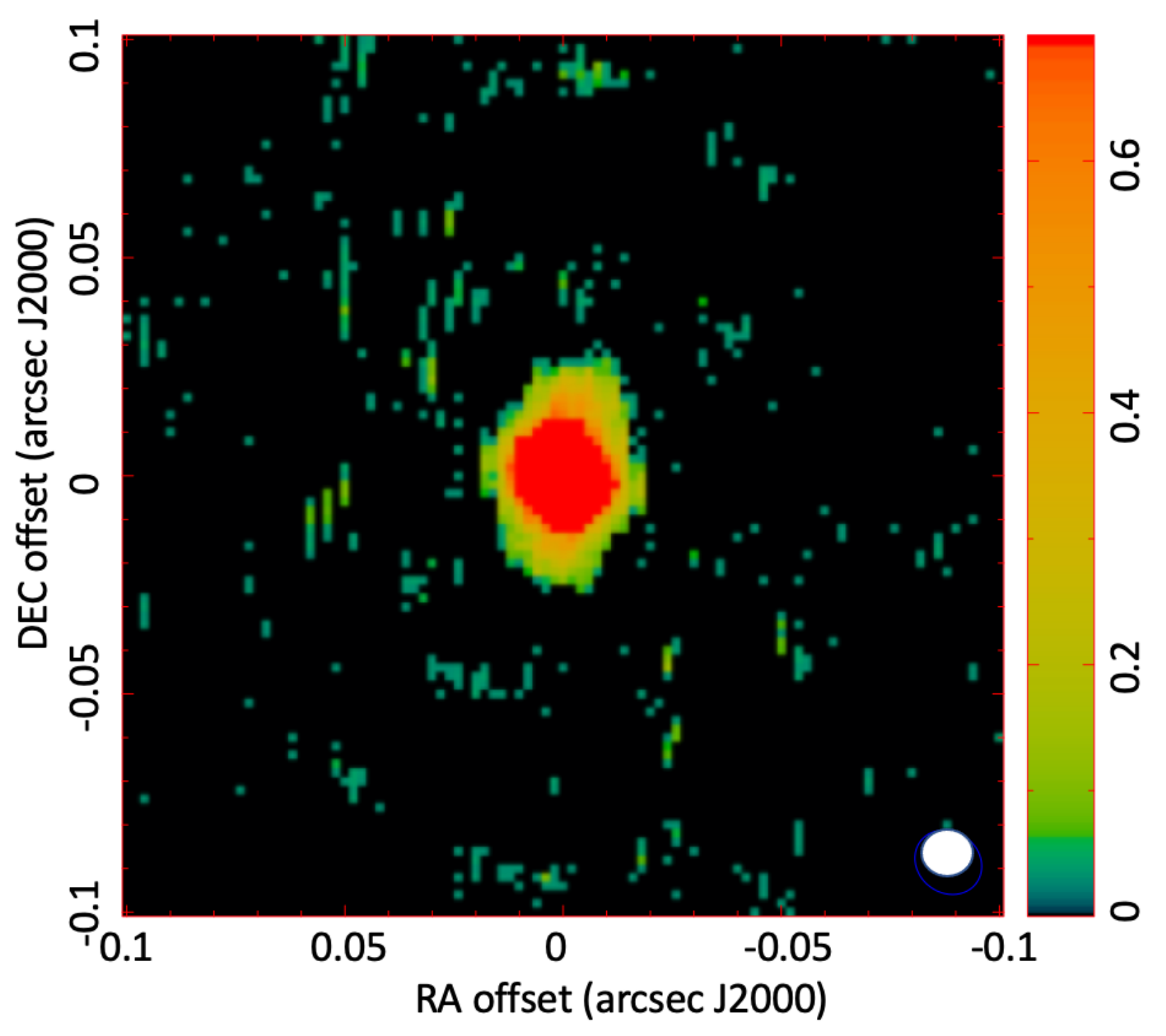}
\includegraphics[width=0.72\columnwidth,angle=0]{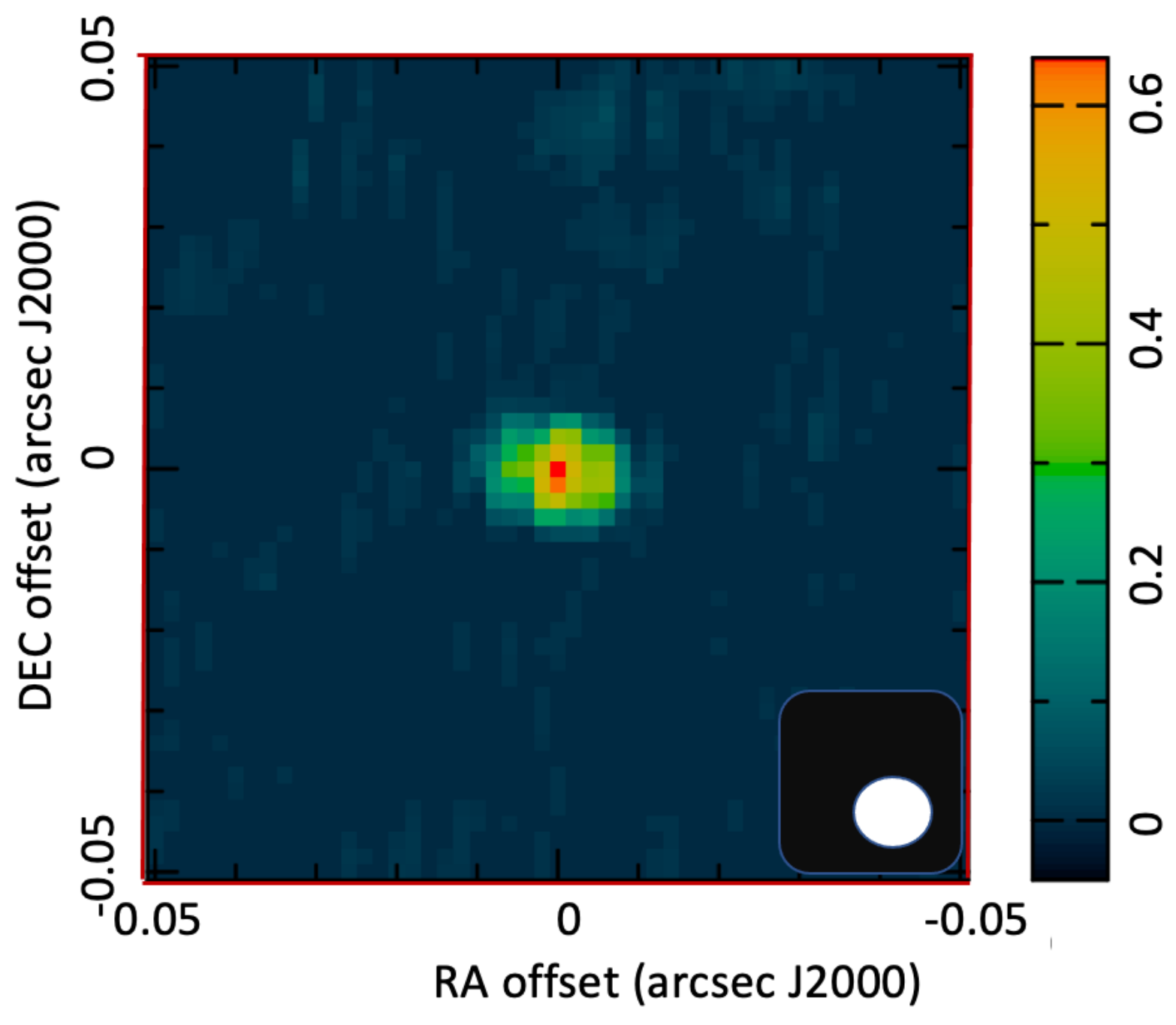} 
\includegraphics[width=0.7\columnwidth,angle=0]{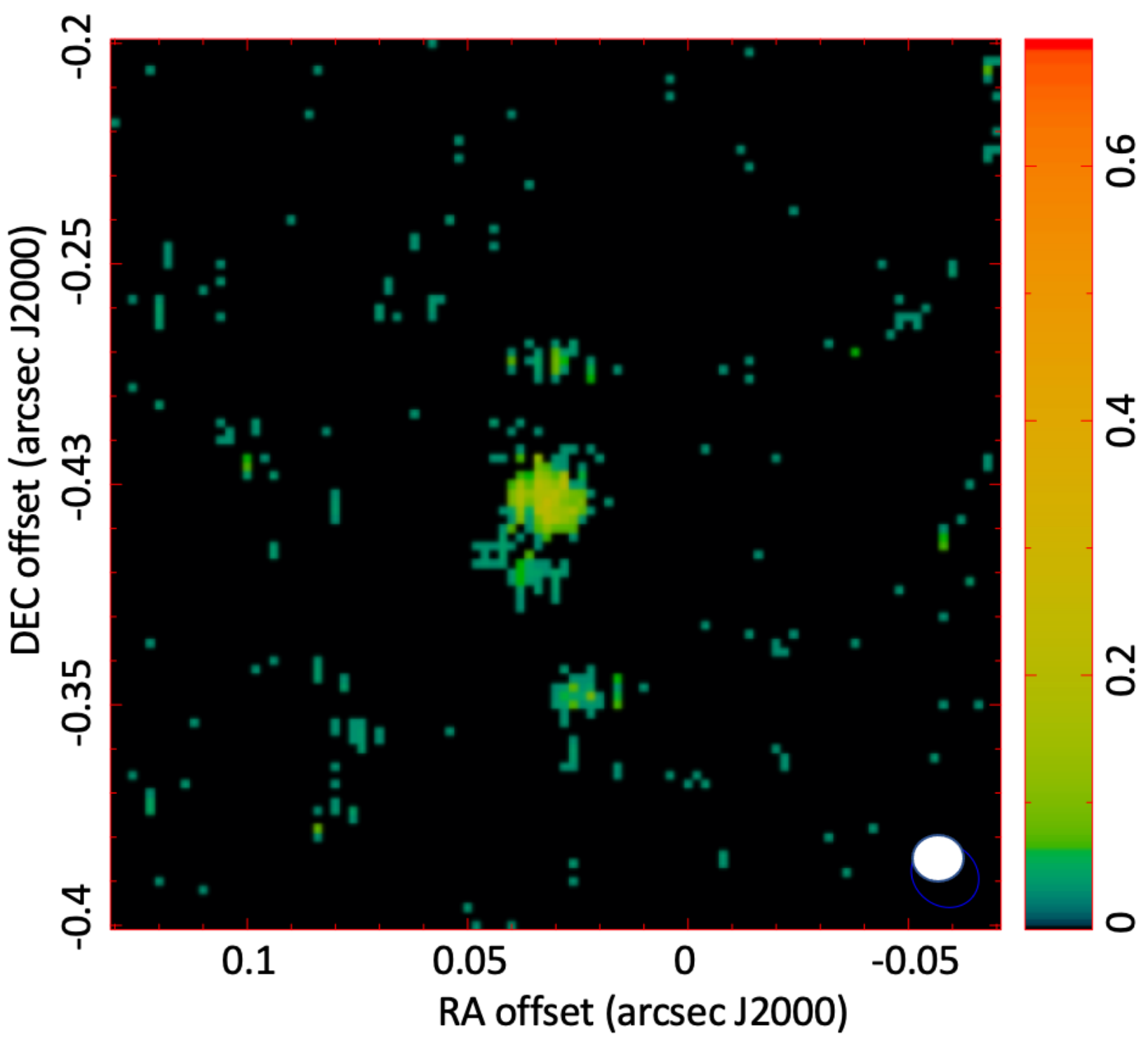}
\includegraphics[width=0.7\columnwidth,angle=0]{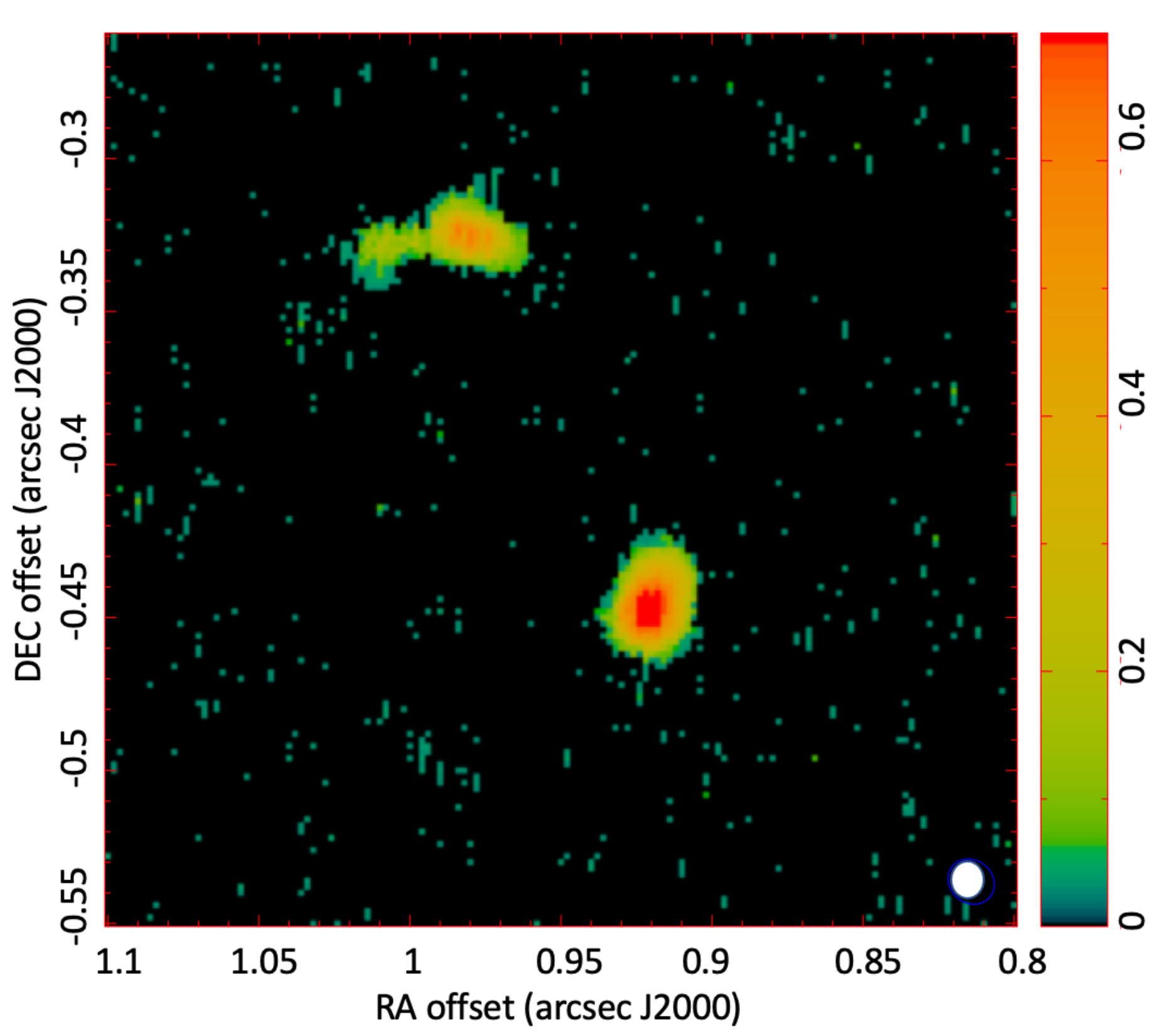}
\caption{Upper panel: Velocity-integrated intensity 0Moment maps of the OH 1667 and 1665 MHz line emission from the EVN data. 
Top: W1 component at 1667 MHz; 
Top Bottom: W1 component at 1665 MHz (see also Fig. \ref{fig9}; 
Bottom Top Right: W2 component at 1667 MHz and located 320 mas below W1;
Bottom: E1 and E2 components at 1667 MHz separated by 126 mas.
The beam size for these images is 13 x 11 mas.
The colour scale is logarithmic in units of Jy beam$^{-1}$.
}
\label{fig8}
\end{center}
\end{figure}

The 1667 MHz emission features in the lower resolution MERLIN data cover a broad velocity range of 5200$-$5600 \kms and encompasses the suspected optical velocities of the West nucleus of about 5360 \kms and of the East nucleus of 5425 \kms (Fig. \ref{fig3}). 
These broad profiles also suggest that the bulk of the 1667 MHz OH emission in Arp\,220 originates in more extended regions at both nuclei and shows little detail about the underlying nuclei themselves.

The velocity gradient (1Moment) map and the velocity width (2Moment) map of the 1667 MHz line emission at both nuclei are presented in Figure \ref{fig4} using a 3$\sigma$ flux cutoff. 
Arp\,220\,E shows a positive velocity gradient in a northeast direction, which suggests rotation within the nuclear regions.  
The Arp\,220\,W nuclear region and the west extension component suggest a very weak but continuous gradient towards the north.
The linewidth distribution of the East component depicted in Figure \ref{fig4} appears rather uniform, while those of the West component are dramatically different with linewidths up 80 \kms in the South component below the core and as low as 20 \kms in the North component.

\begin{figure}
\begin{center}
\includegraphics[width=1.0\columnwidth,angle=0]{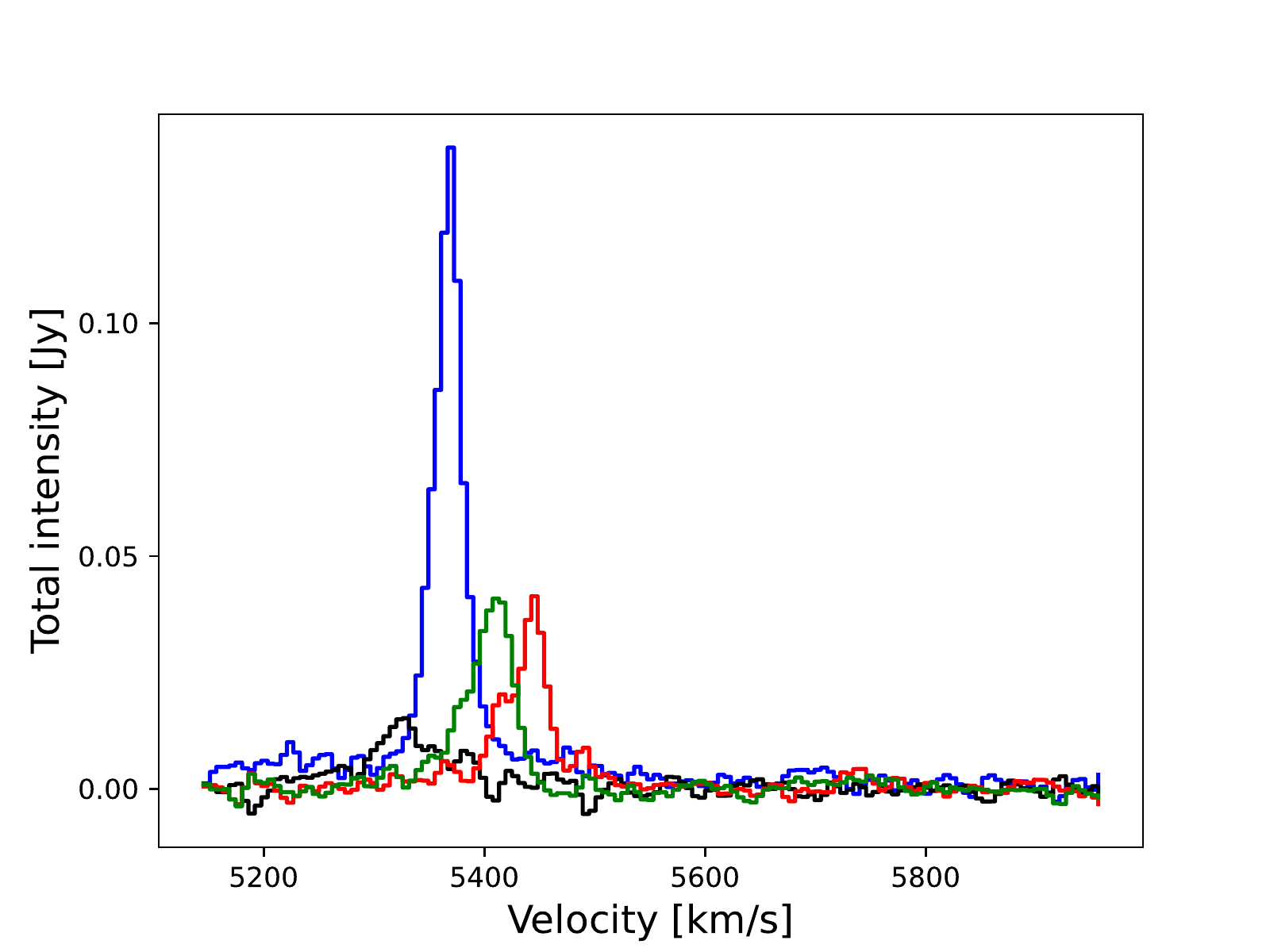}
\caption{Spectra of the 1667 MHz OH emission from the EVN data.
The spectra are for W1-N(blue), W2-S(black), double feature E1-E(red) and E2-S(green).
The zero level base of these spectra covers about 300 \kms from 5250 to 5550 \kms with the W1 spectrum showing a plateaux at about 10 mJy reaching further down in velocity.
The W1 spectrum also shows the weak 1665 MHz emission at about 5700 \kmss.
Velocities using an optical definition and the line widths are presented in Table \ref{tab3}.}
\label{fig9}
\end{center}
\end{figure}

\begin{figure*}
\begin{center}
\includegraphics[width=1.6\columnwidth,angle=0]{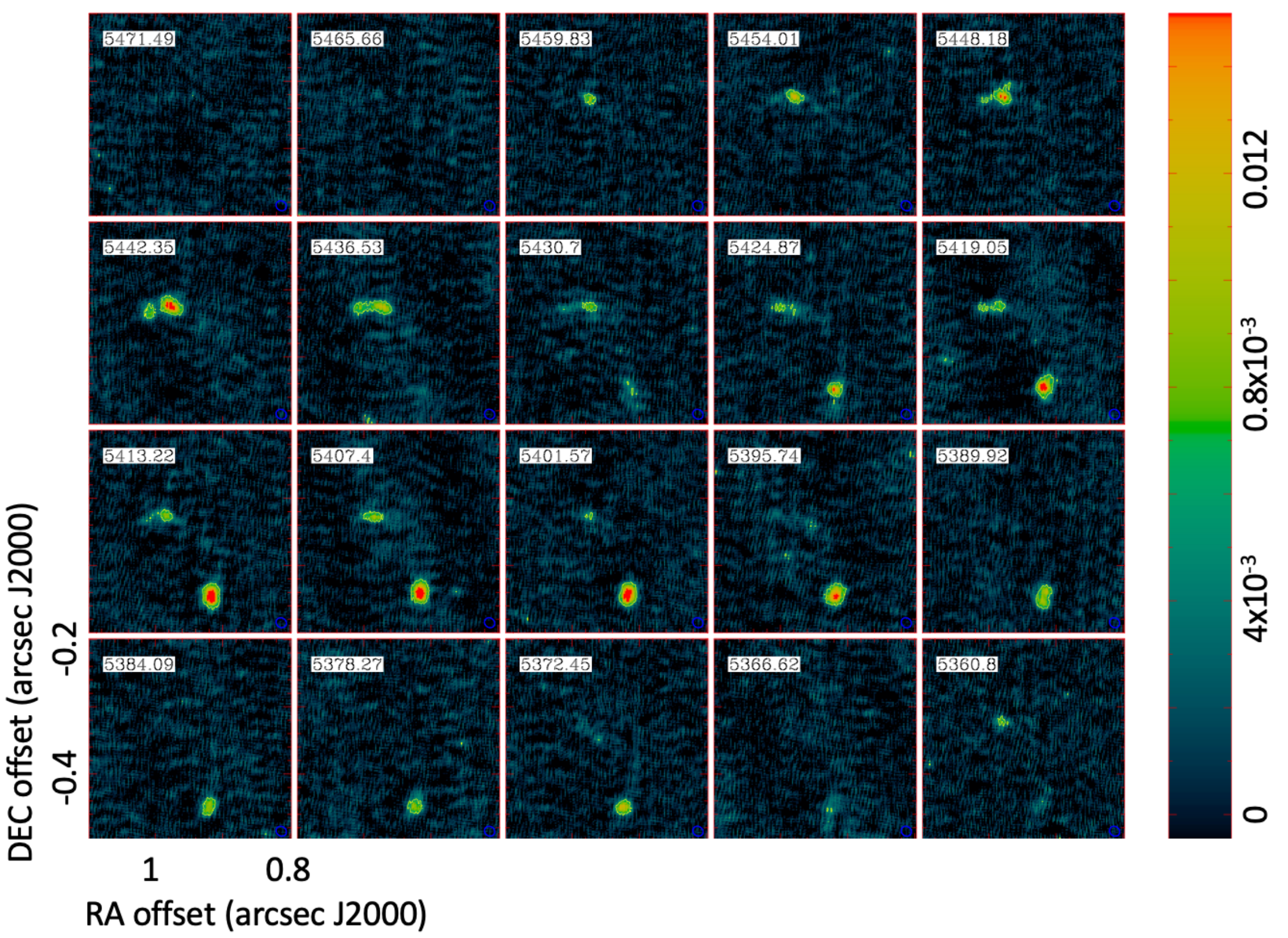}
\caption{Channel maps of OH 1667 line emission in the Arp\,220E from the EVN data. 
Two emission regions are identified as the extended and filamentary region E1 in the north and the compact region E2 in the south. 
Some additional weak regions may be found between these two regions.
The restoring beam is 0.16$\times$0.14 mas$^2$ at PA=53$\degr$. 
The rms noise level is 1.3 mJy beam$^{-1}$. 
The peak intensity of 19.8 mJy beam$^{-1}$ is found for E2 in the 5407.4 \kms channel. 
For comparison, the peak intensity of the 1667 MHz line is 135 mJy beam$^{-1}$ in the 5366.6 \kms channel. 
For image clarity, only the 3$\sigma$ contour of 5.2 mJy beam$^{-1}$ is plotted.
The velocity axis is based on the optical definition of velocity and the channel width is 5.8 \kmss.}
\label{fig10}
\end{center}
\end{figure*}

\subsection{The MERLIN 1665 MHz OH line emission}
\label{sec33}

The  new 1665 MHz OH line cube also covers a velocity range of 5600$-$5900 \kms with a velocity resolution of 11.7 \kmss. 
The 1665 MHz channel maps and the 0Moment and 1Moment maps are presented in Figures \ref{fig5} and \ref{fig6}.
Contrary to the observed emission structure of the 1667 MHz line, the observable/prominent 1665 MHz emission appears only at the components Arp\,220E and Arp\,220W-South, as may also be deduced from the spectra of Figure \ref{fig3}. 
Any difference between the 1667 and 1665 MHz emission structures may result from a different dynamic range in Figure \ref{fig5} and \ref{fig6}.

The west-east elongated 1665 MHz emission component in Arp\,220E shows a continuous velocity gradient, which consistent with the 1667 MHz line findings (Fig. \ref{fig3}). 
An overall hyperfine line ratio $R_H(67/65)$ = 3.6 would correspond to an amplifying optical depth $\tau_{67}$ = 2.4 (see Sect. \ref{sec5}).  

Arp\,220W-South shows a broad 1665 MHz spectrum possibly made up of multiple velocity components (Fig. \ref{fig3}).
The line ratio $R_H(67/65)$ = 5.0 suggests a higher amplifying optical depth $\tau_{67}$ = 3.3 but there is no clear 1665 MHz velocity gradient. 
Surprisingly, Arp\,220W-North shows no clear 1665 MHz emission and its narrow 1667 MHz emission component appears to have a large amplifying optical depth $\tau_{67}$ and a possible small gradient visible in the 1Moment map (Fig. \ref{fig6}).
This narrow feature is associated with the compact source found in the EVN data (see Fig. \ref{fig9} and Sect. \ref{sec35} below).

\subsection{EVN imaging of the continuum emission}
\label{sec34}

The continuum structure of Arp\,220 at high-resolution is known to contain a number of identifiable supernova remnants (SNRs) at the two nuclei \citep{SmithEA1998,LonsdaleEA2006,ParraEA2007,VareniusEA2019}. 
The new EVN image shows two groups of point sources covering the East and West nuclear regions consistent with the continuum peaks (the crosses) in the MERLIN data in Figure \ref{fig1}.
Although some side-lobe artefacts, phase errors, and RFI-related stripes may still be present in the map, the general configuration of the point sources is consistent with earlier detections even if their flux densities and positions appear not fully consistent with earlier experiments  \citep{SmithEA1998,LonsdaleEA2006,ParraEA2007}. 
Our sources are found to be brighter than earlier detections but variations in flux and position are to be expected in an evolving starburst environment.
The presence of the point sources in the map of Figure \ref{fig7} accounts for about 90 mJy beam$^{-1}$, which is about 40\% of the integrated flux of the East and West radio nuclei.
A thorough analysis still needs to be made of the power spectra and locations of these point sources and their (re-)appearance in comparison with earlier experiments.

The configuration of the SNR point sources confirms the star formation nature of the nuclear regions.
The slightly elliptical N-S source configuration appears centred in between the two OH emission regions and coincides with the H$_2$CO emission regions \citep{BaanEA2017}.
The West configuration appears to have a N-S absorption lane possibly related to the edge-on torus at this nucleus (see Section \ref{sec43})
It should be noted that the point sources at the Eastern nucleus appear less dense and elongated in the SW-NE direction forming a connection with the continuum bridge below and between the two nuclei.

\subsection{EVN Imaging of the OH Line Emission}
\label{sec35}

The new high-resolution EVN observations provides more details about and confirm the existence of four compact high-brightness VLBI maser components in Arp\,220, two associated with each of the nuclear regions \citep{LonsdaleEA1998,RovilosEA2003}. 
Following the nomenclature based on earlier VLBI detections, they have been named W1 (north), W2 (south), E1 (east) and E2 (south). 
The velocity-integrated maps of these four features are presented in Figure \ref{fig8} and the integrated spectra are shown in Figure \ref{fig9}. 
Together they account for about 15\% of the total line emission in Arp\,220.

The most prominent emission feature W1 is located at the southeastern edge of the Arp\,220W-North and appears as a point source in channel maps across an optical velocity range 5416$-$5498 \kms with a peak at about 5364 \kmss. 
W1 also shows compact emission at a much lower velocity of 5360 \kms, which may result from foreground emission (see Sect. \ref{sec43}).
The newly detected 1665 MHz counterpart of W1 (Fig. \ref{fig8}(top right) shows a point-like centre and a mysterious E-W (halo or scattering) extension and peaks at  the velocity of 5586 \kmss. 
The less compact W2 feature is only detected at 1667 MHz and appears with two weak companions at the southern edge of the South region at about 115 pc (0\farcs3) south of W1. 

The compact components E1 (east) and E2 (south) at the Eastern nucleus are shown in the channel maps of Figure \ref{fig10}.
Compared with W1, the elongated E1 component is redshifted in the range 5401$-$5461 \kms with a primary peak at 5443 \kms at the brighter head and a secondary peak at 5399 \kms at the (slightly redshifted) eastern tail. 
The compact component E2 is located (0\farcs14) 49 pc southwest of E1 and is also redshifted relative to W1 with a peak at 5409 \kmss.
E2 shows a northwest extension resulting from a possible double structure and is more prominent than was found in earlier experiments \citep{LonsdaleEA1998}.
The peaks of the two E features confrim a large-scale northeast velocity gradient along the structure of the East nucleus. 
In addition to E1 and E2, several other compact weak point sources may be identified in the field.

The relative positions of the four components has been preserved during self-calibration (see Section \ref{sec22}), their actual locations within the East and West nuclei will be evident as they re-appear in the combined EVN - MERLIN emission maps presented below (see Section \ref{sec4}).
The spectral characteristics of the EVN-detected features are also presented in Table \ref{tab3}.

\begin{figure}
\begin{center}
\includegraphics[width=0.9\columnwidth,angle=-90]{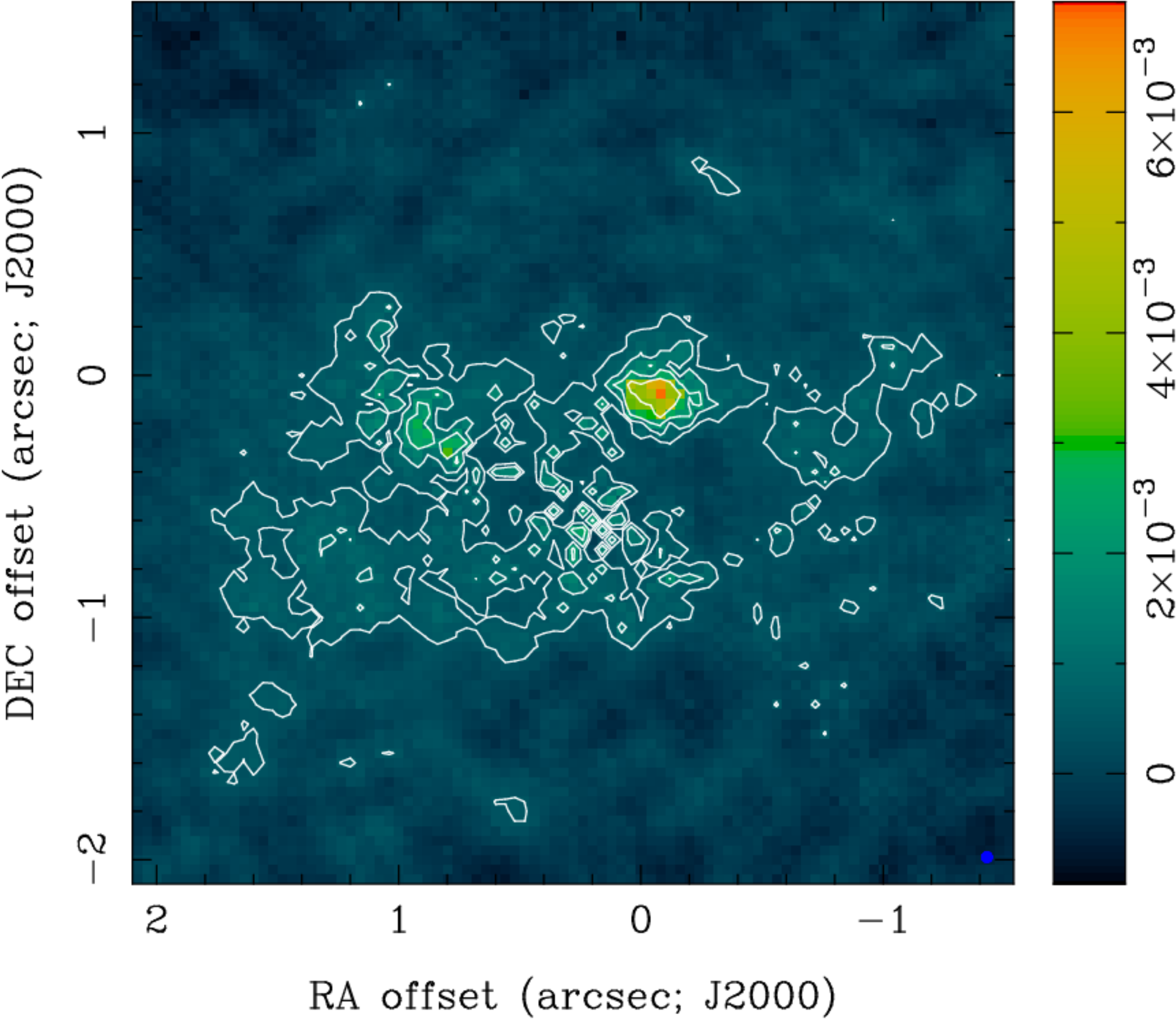}
\caption{The 1.6-GHz continuum emission of Arp\,220 made from the combined MERLIN-EVN data.
The restoring beam is 0\farcs04 $\times$ 0\farcs04, PA=74\fdg3. 
The peak intensity is 6.6 mJy beam$^{-1}$, and the rms noise in the off-source region is 0.24 mJy beam$^{-1}$. 
The contours are at 0.48 mJy beam$^{-1}$ $\times$(1, 2, 4, and 8).}
The intensity colour scale is in units of mJy beam$^{-1}$.
\label{fig11}
\end{center}
\end{figure}

\begin{figure}
\begin{center}
\includegraphics[width=1.0\columnwidth,angle=0]{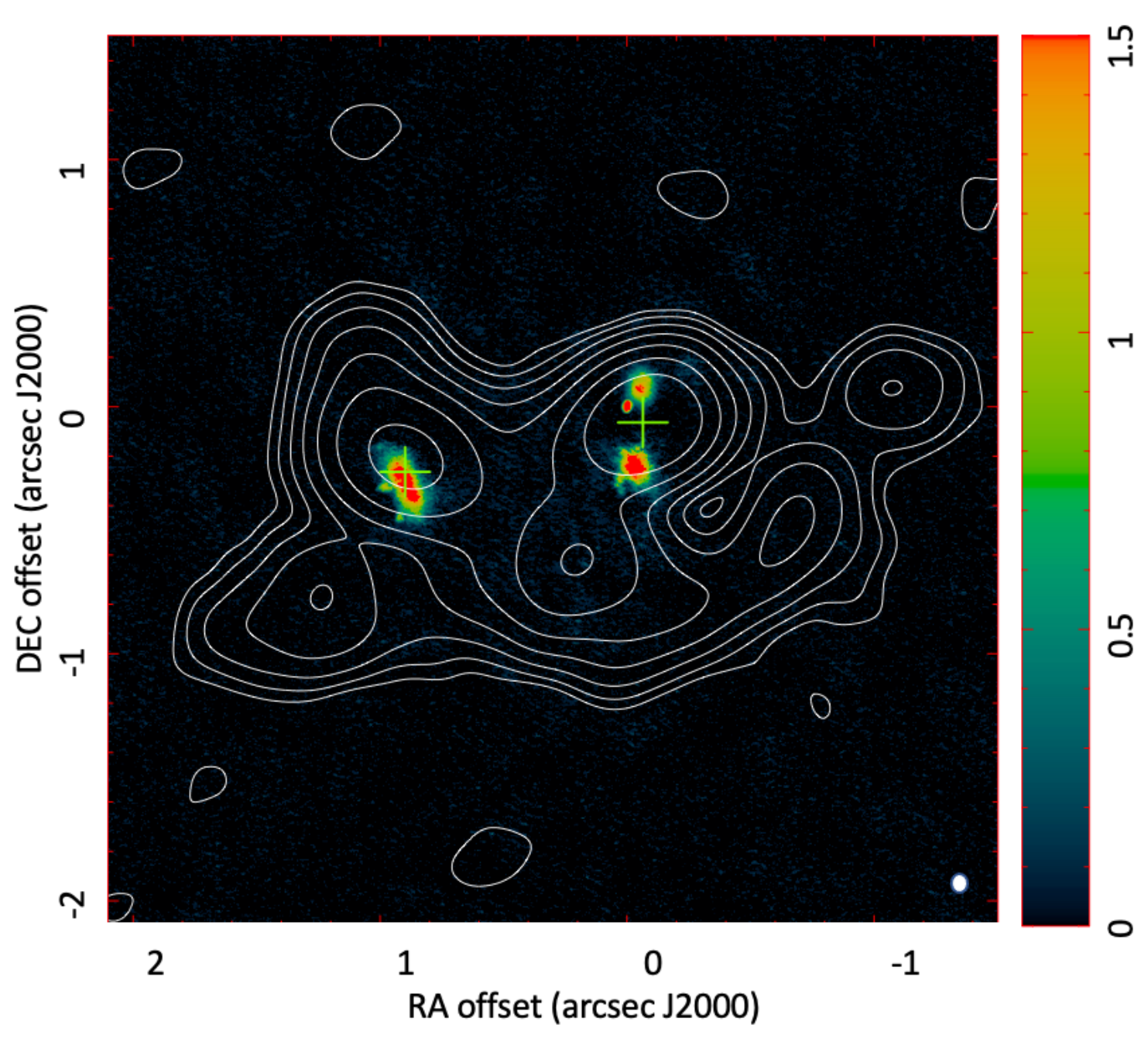}
\caption{Velocity-integrated intensity maps of the OH 1667 MHz emission using the combined MERLIN - EVN data. 
The Integrated 1667 MHz OH map overlaid on the MERLIN 1.6 GHz continuum.
The continuum contours of the MERLIN continuum are at 0.80 mJy beam$^{-1}$ $\times$(1, 2, 4, 8, 16, 32, 64).
The intensity colour scale is logarithmic in units of mJy beam$^{-1}$.
The peak integrated intensity in the map is 93 mJy beam$^{-1}$.
The two crosses represent the dust emission peak positions at 230 GHz.
The scale size of Figure \ref{fig11} and this Figure is identical.}
\label{fig12}
\end{center}
\end{figure}

\section{Combining MERLIN and EVN data}\label{sec4} 

\subsection{The MERLIN-EVN imaging of Arp\,220}
  \label{sec41}
  
The continuum structure with the two nuclear components obtained from the combined ME data shows again that the Western nucleus is more compact than the Eastern nucleus and shows the extended structure below the nuclei found in the MERLIN data alone (Fig. \ref{fig11}).

\begin{figure}
\begin{center}
\vspace{-2mm}
\includegraphics[width=0.7\columnwidth,angle=-90]{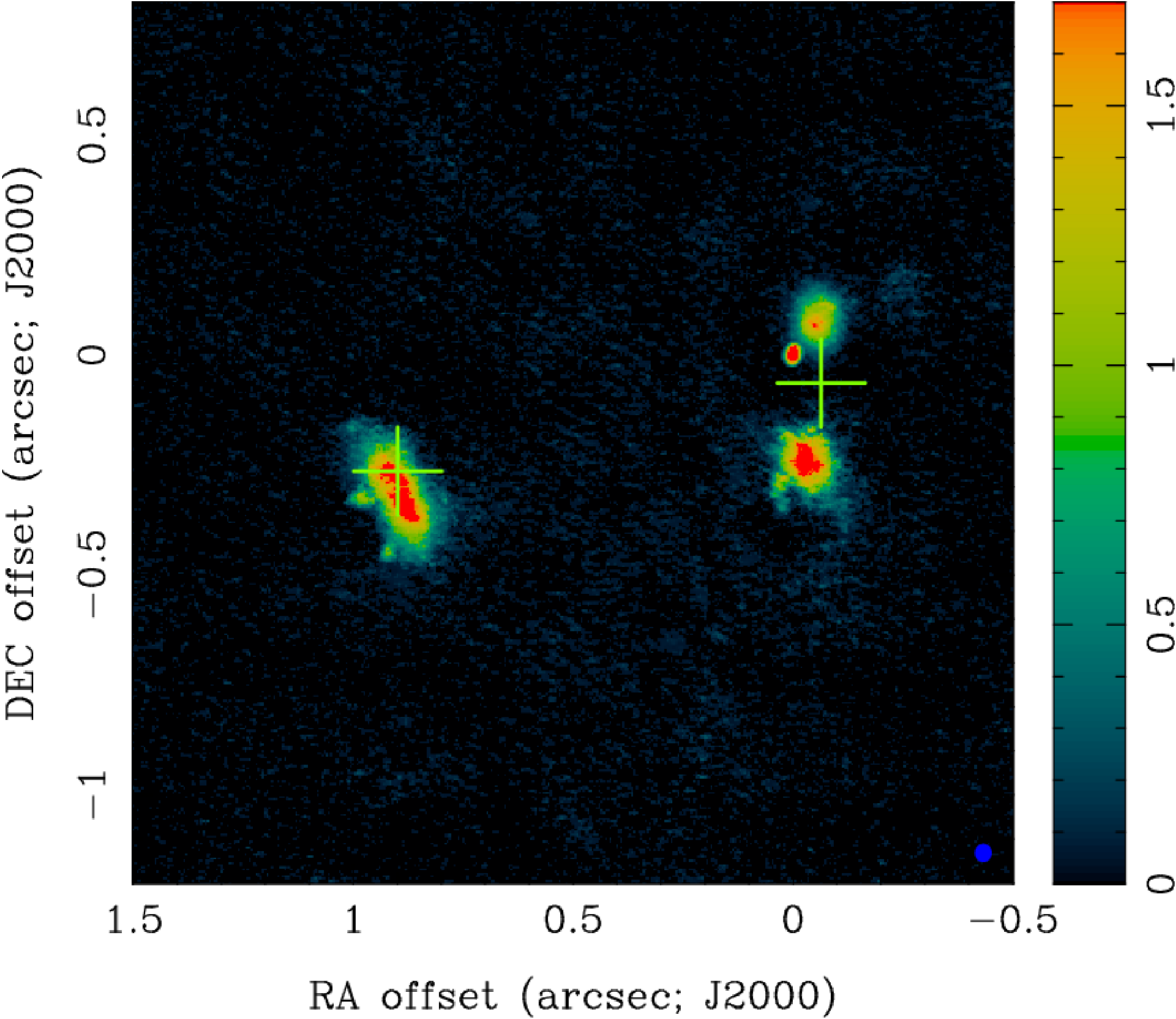}
\vspace{0mm}
\includegraphics[width=0.7\columnwidth,angle=-90]{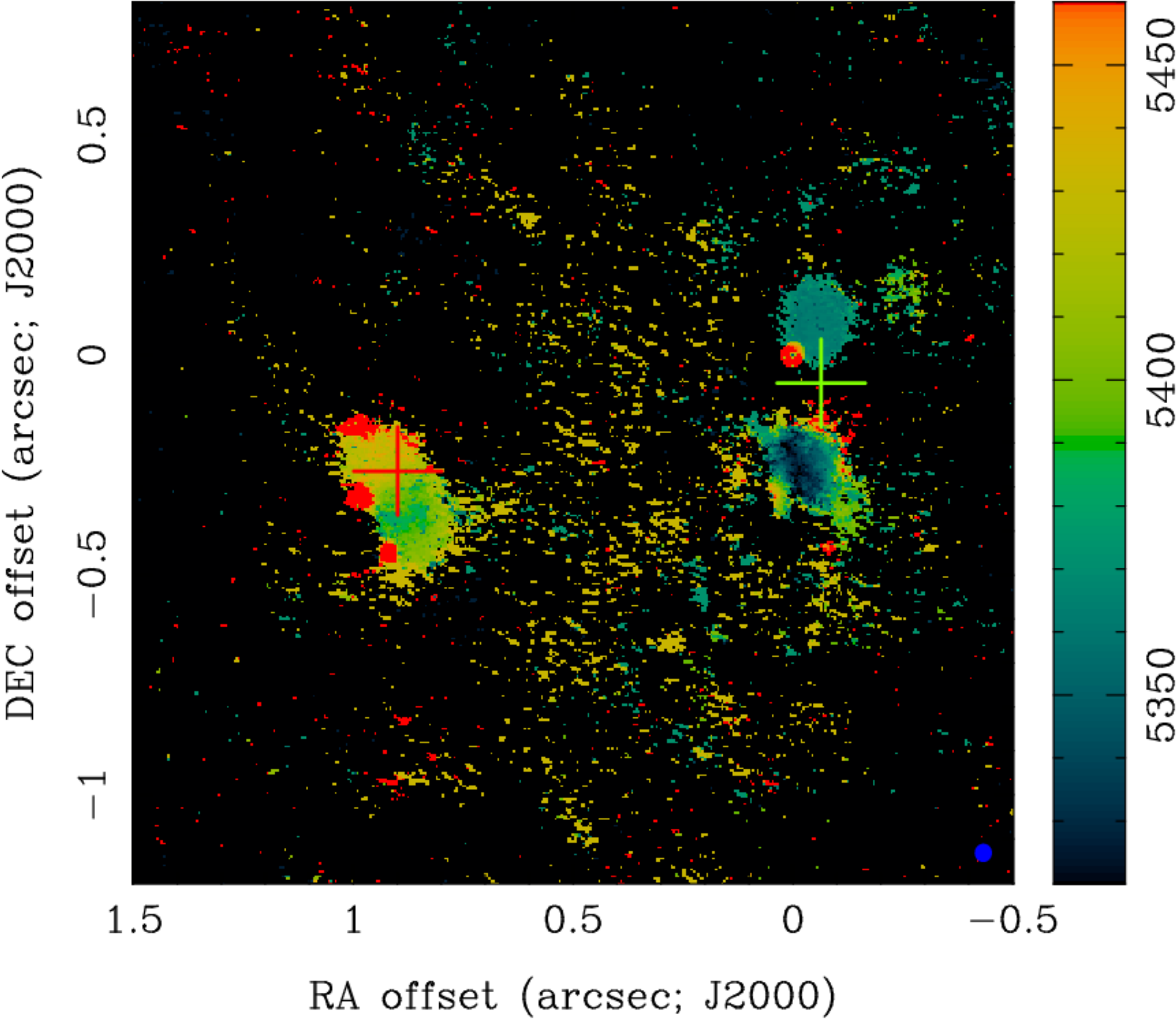}
\vspace{0mm}
\includegraphics[width=0.7\columnwidth,angle=-90]{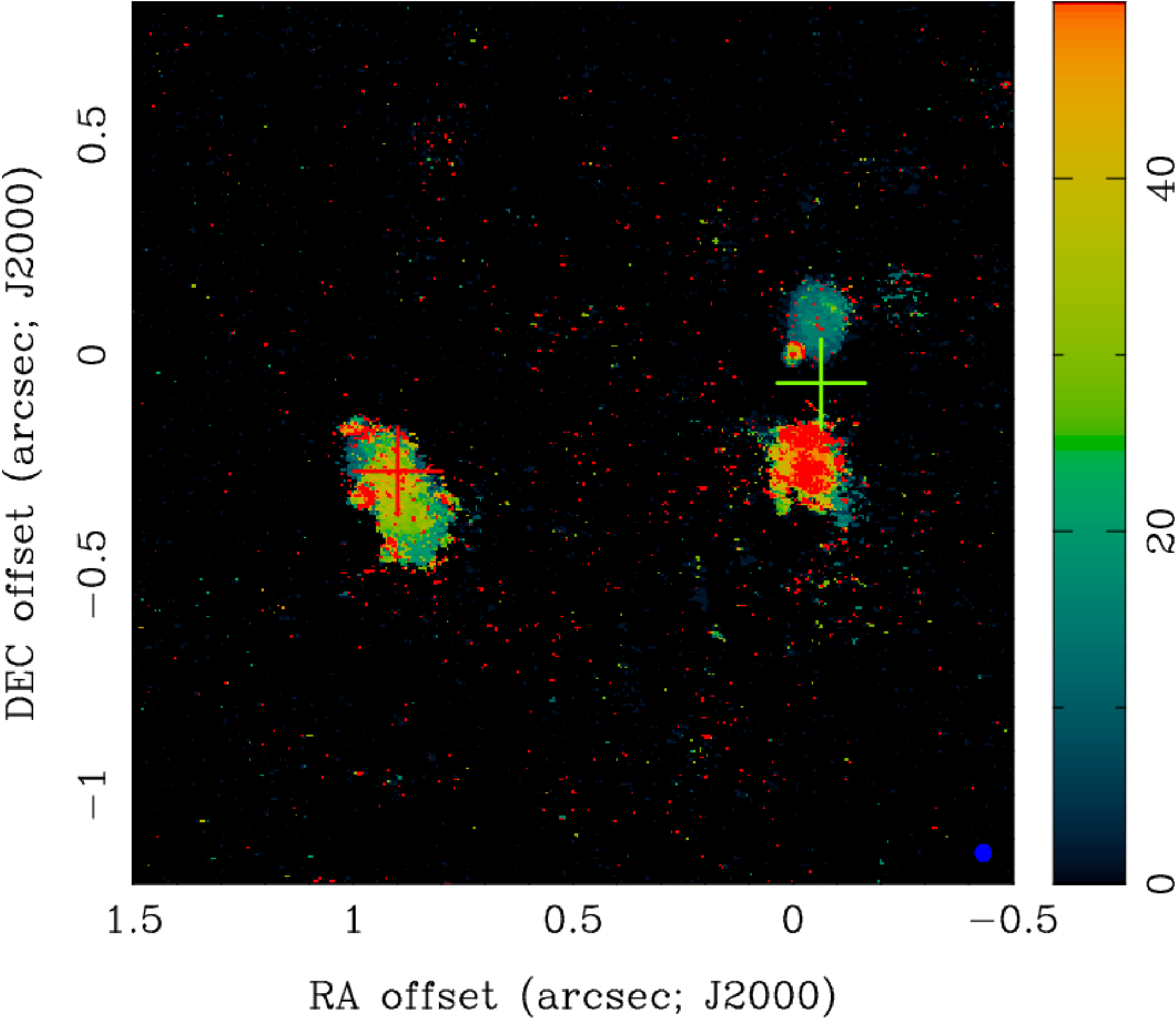}
\vspace{3mm}
\caption{The combined MERLIN-EVN data of the 1667 MHz OH emission. 
(Top) The Moment 0 velocity-integrated intensity map of OH 1667 emission. 
The restoring beam size is 40 $\times$ 40 mas.
Four of the six compact components seen at the two nuclei represent components identified within the EVN data (see Fig. \ref{fig8}).
The integrated intensity colour scale is logarithmic in units of Jy beam$^{-1}$.
The peak values are for the 
East component   18.15 Jy beam$^{-1}$,
the West-South component 11.25 Jy beam$^{-1}$, and
the West-North component 7.44 Jy beam$^{-1}$ with an estimated error of 0.011 Jy beam$^{-1}$.
(Middle) The Moment 1 velocity map of the OH 1667 emission.
The colour bar indicates the radial velocity in \kmss.
(Bottom) The Moment 2 velocity width map of the OH 1667 emission.
The colour bar indicates the velocity widths in \kmss.}
\label{fig13}
\end{center}
\end{figure}

\begin{figure}
\begin{center}
\vspace{0mm}
\includegraphics[width=0.7\columnwidth,angle=-90]{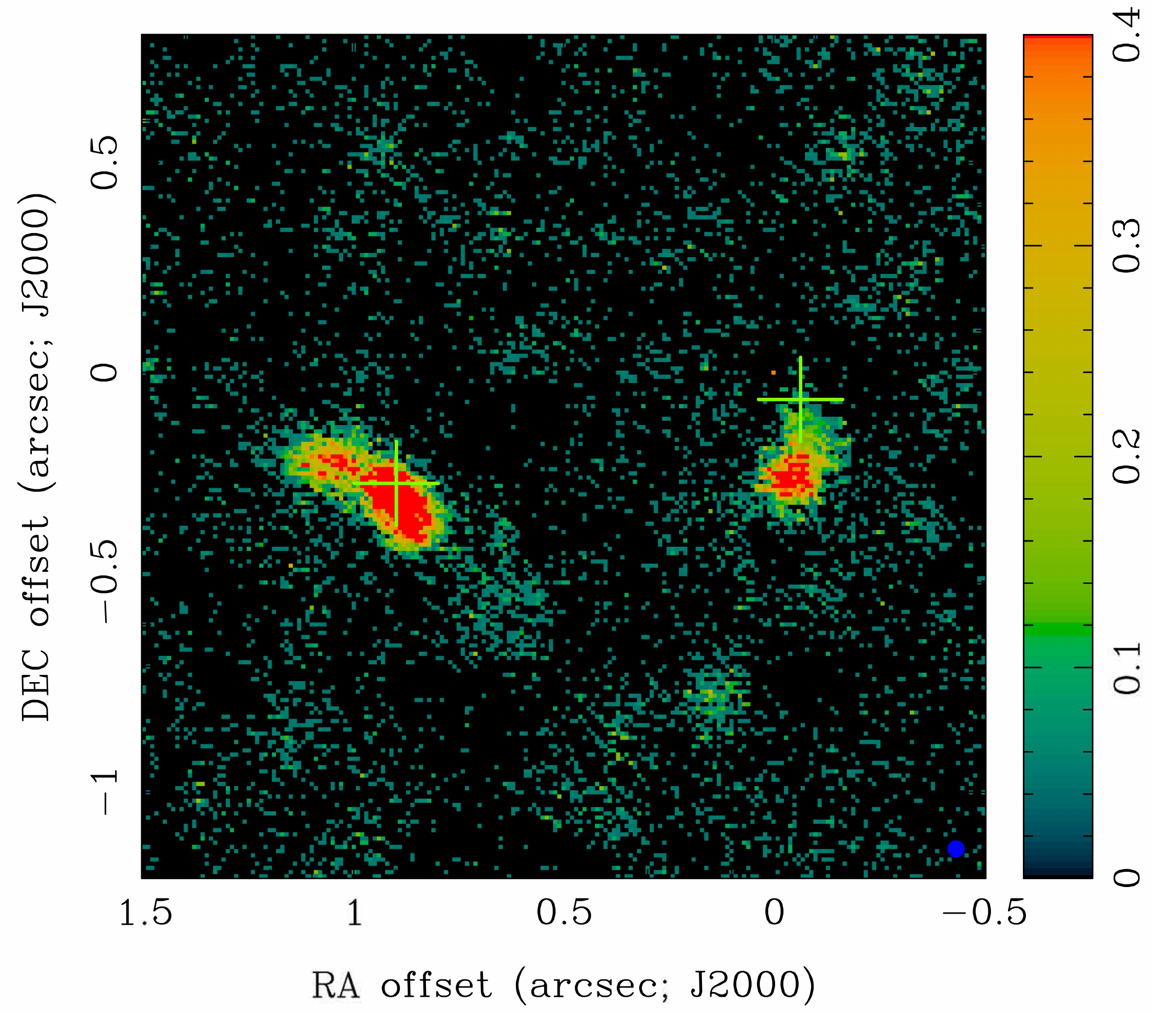}
\vspace{0mm}
\includegraphics[width=0.7\columnwidth,angle=-90]{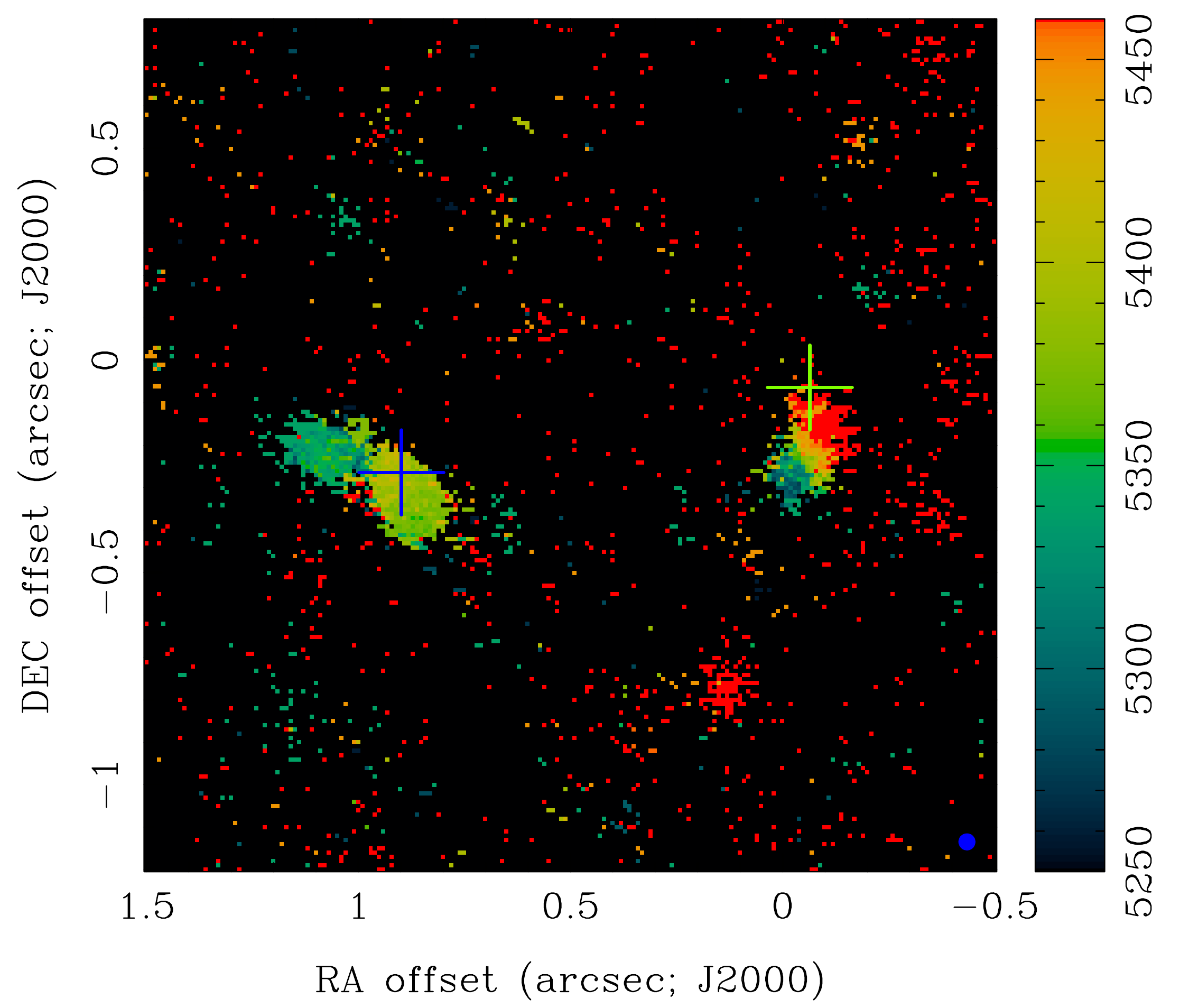}
\vspace{0mm}
\includegraphics[width=0.7\columnwidth,angle=-90]{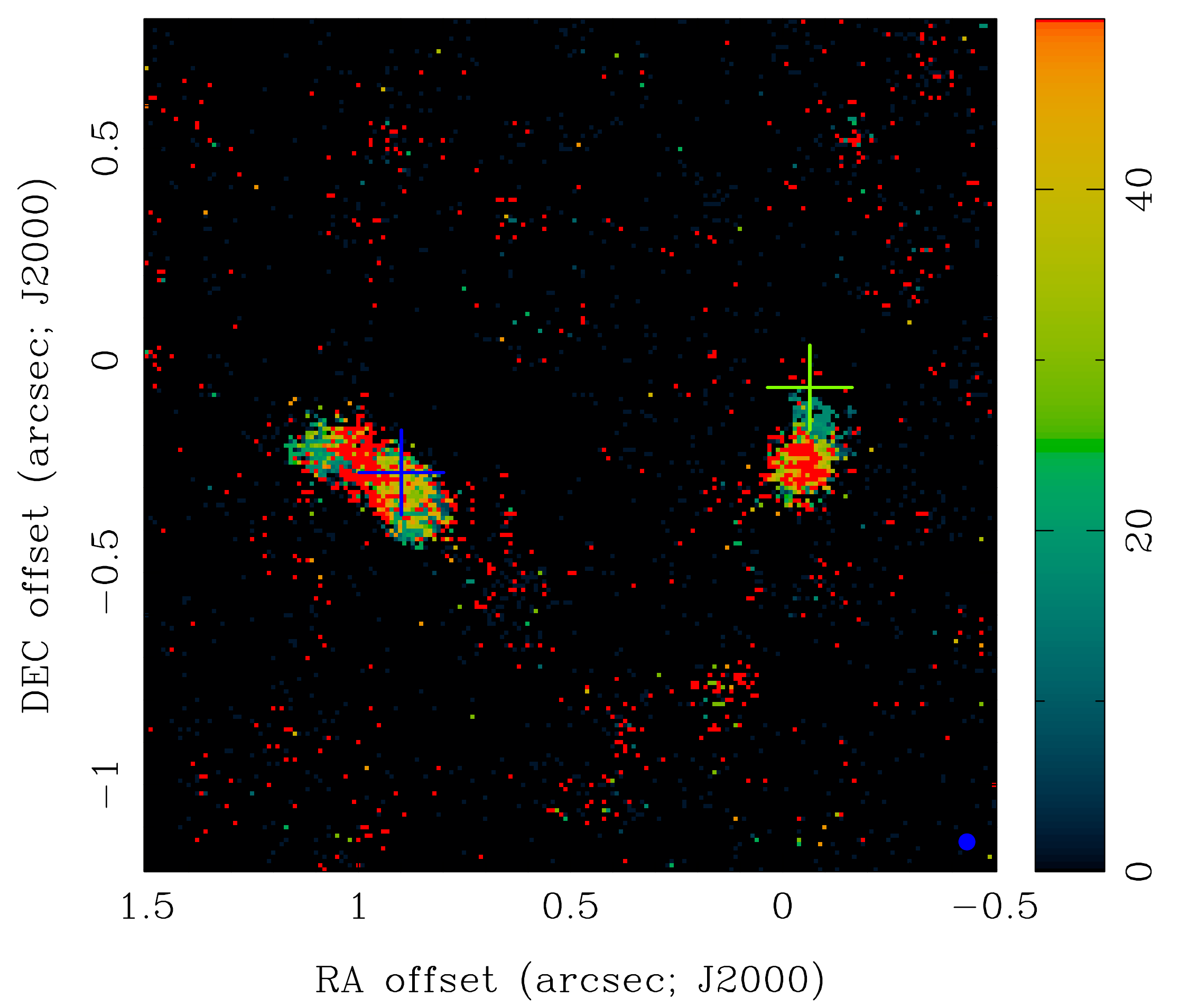}
\vspace{3mm}
\caption{The combined MERLIN-EVN data of the 1665 MHz OH emission. 
(Top) The Moment 0 velocity-integrated intensity map of the 1665 MHz OH emission. 
The restoring beam size is 40 mas.
The arrows in the map indicate the positions of the four VLBI components identified within the EVN data.
The unit of the logarithmic colour scale is Jy beam$^{-1}$
The integrated flux densities of the components are for the 
East component and  West-South component 4.0  Jy beam$^{-1}$, and 
the West-North component 0.1 Jy beam$^{-1}$ with estimate error of 0.011 Jy beam$^{-1}$.
(Middle) The Moment 1 velocity map of the OH 1665 emission.
The colour bar indicates the radial velocity in \kmss.
(Bottom) The Moment 2 velocity width map of the OH 1665 emission.
The colour bar indicates the velocity widths in \kmss.}
\label{fig14}
\end{center}
\end{figure}

The extended and compact OH emission components have been presented in the images of Figures \ref{fig12}, \ref{fig13} and \ref{fig14}.
The extended 1667 MHz and 1665 MHz emission regions cover the two nuclear regions and have a linear extent of  80 and 106 pc for Arp\,220W North and South and 154 pc for Arp\,220E.
While the OH emission region at the East nucleus falls slightly below the continuum peak, the two regions at the West nucleus straddle the continuum peak, except that the North 1665 MHz emission in Arp\.220W appears much weaker.
Any further apparent structural differences in the 1667 and 1665 MHz images between North and South in West nucleus the eastern part of the East nucleus in Figs. \ref{fig13} and \ref{fig14} result from a difference in dynamic range of the two 0Moment maps.
 
Adopting a line-of-sight maser amplification scenario for the OH maser emission, the OH emitting regions and the far-infrared pumping source and the amplified continuum must indeed be on the line-of-sight.
As a result, any amplified OH emission regions are always superposed on the continuum structure and the  FIR radiation fields of the source. 

The relative locations of the OH emission regions may follow from the line-of-sight velocity information at the two nuclei and the velocity fields of other molecular constituents of the Arp\,220 system as seen in the recent  $^{12}$CO\,(3 - 2) and $^{13}$CO\,(4 - 3) observations \citep{WheelerEA2020} and earlier observations of $^{12}$CO\,(3 - 2) \citep{SakamotoEA1999,SakamotoEA2008} and $^{12}$CO\,(6 - 5) emission \citep{RangwalaEA2015}.
The $^{12}$CO\,(3 - 2) emission structure in Arp\,220 shows that the two nuclei are embedded in (or covered by) a diffuse and drifting molecular structure encompassing a region comparable to the extended kpc-sized radio continuum structure shown in Figure \ref{fig1}.
Therefore, excited OH molecules within the drifting molecular foreground may contribute to the amplification of the background continuum and re-amplification of the systemic emissions originating in the merging galaxies. 
 
\begin{figure}
\begin{center}
\vspace{-1mm}
\includegraphics[width=1.02\columnwidth,angle=0]{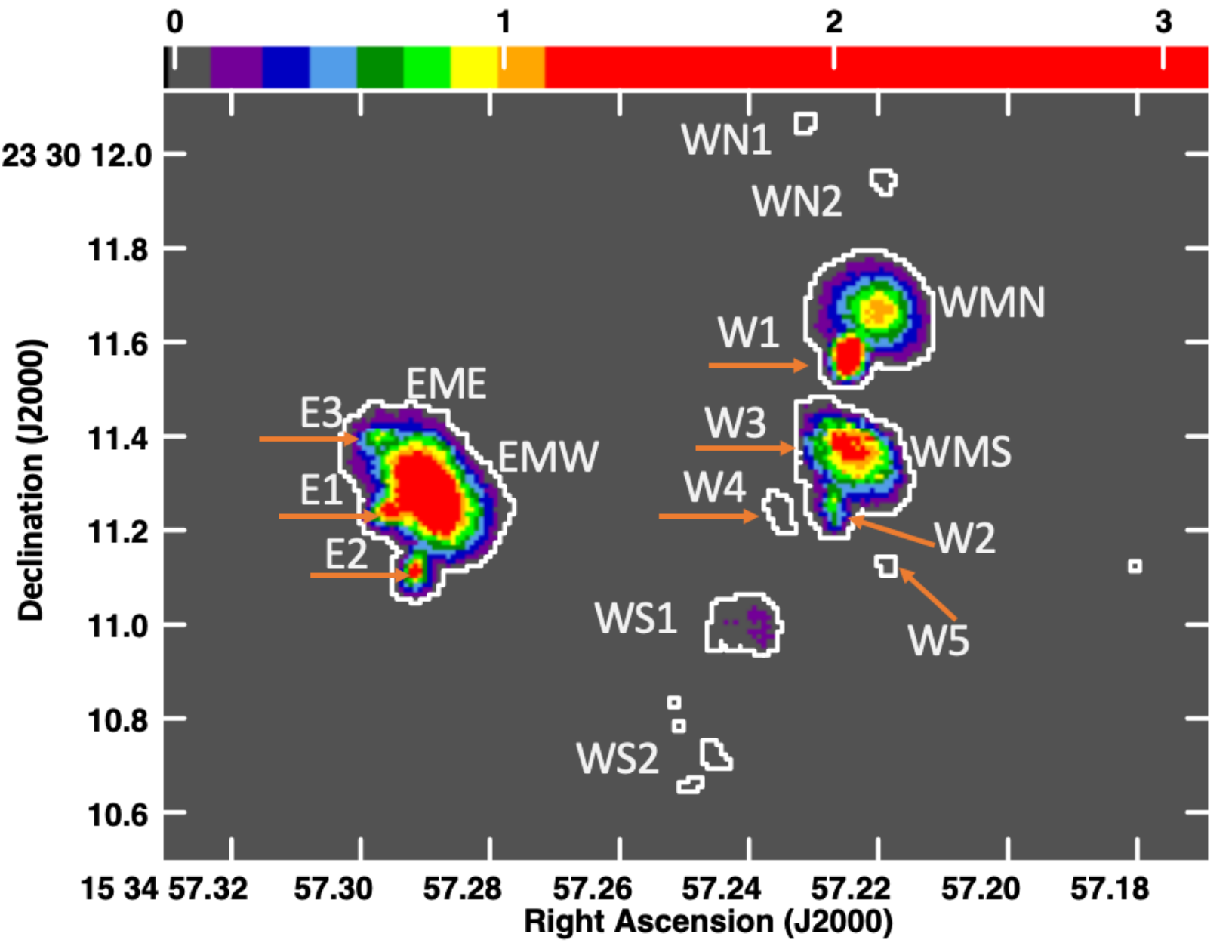}
\caption{A deep Moment Zero map of the 1667 MHz OH emission in the extended region around 
the two nuclei based on the combined MERLIN - EVN data.
The contours have been set at 0.2 mJy beam$^{-1}$, which is 1.5 times the rms in the map.
The designations of all identified compact and extended regions have been indicated in reference to the spectral information presented in Table \ref{tab3} and Figures \ref{fig16} and \ref{fig17}.
The spatial scale of the diagram of 0\farcs1 corresponds to 38 pc.
All observed emission components lie within the continuum contours of the two galaxies as presented in Figures \ref{fig11} and \ref{fig12}.}
\label{fig15}
\end{center}
\end{figure}

\begin{figure*}
\begin{center}
\includegraphics[width=2.1\columnwidth,angle=0]{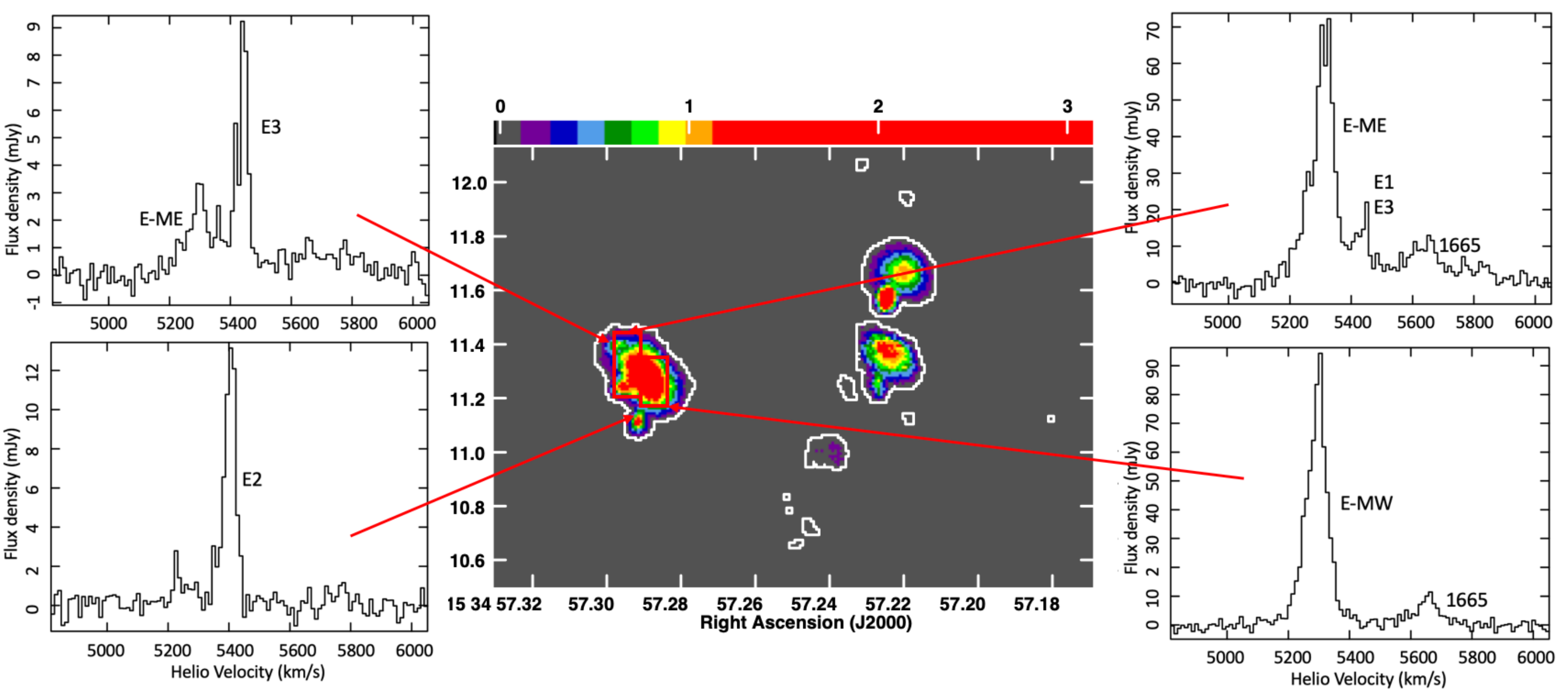} \\
\caption{Spectral components in Arp\,220E.
The velocity range of all spectra runs from 4809 to 6046 \kms increasing from left to right.
The vertical scale of all spectra is in mJy except that this scale is not equal for all spectra in order to show the weaker components.}
\label{fig16}
\end{center}
\end{figure*}

\subsection{Extended and compact OH emission in East}\label{sec42}

A deep 0Moment map of the 1667 MHz emission in Arp\,220 shows the dominant emission regions at both the East and West nuclei, in addition to compact emission regions surrounding the two nuclei (Fig. \ref{fig15}).
Three compact emission regions in Arp\,220E appear at the eastern edge of the extended emission regions Arp\,220E, of which two have a counterpart in the EVN data (Fig. \ref{fig8} and \ref{fig10}). 
These compact regions are clearly visible in Figures \ref{fig13} and \ref{fig14} and are identified in Figure \ref{fig15} as E1 - E3. 

The spectral velocity components at all identified  emission regions are displayed in Figures \ref{fig16}.
The three regions E1-E3 at the East nucleus are found to have a velocity close to 5425 \kmss, which is about 110 \kms higher than the velocity of 5314 \kms of the E-MainE and E-MainW extended OH emission regions that appear to be in the foreground.
The observed strength of these regions would result from re-amplification of the maser signature by the extended foreground gas (see Section\,\ref{sec5}).

Composite diagrams with the spectra at all locations in the 0Moment map clearly show evidence for a superposition of structural components with distinct velocity systems at each nucleus in Figures \ref{fig16}. 
A position-velocity diagram in Figure \ref{fig18} depicts these velocity systems using the velocity values obtained from the spectra at various locations (see Table \ref{tab2}).
Assuming that the higher velocities of the compact and surrounding emission components at both nuclei represent the systemic velocities of the two nuclei (or what is left over of them), the lower velocity components must represent foreground gas structures in the system.
This would suggest that the compact components identify a systemic velocity of 5425 \kms for the East nucleus (see Fig. \ref{fig18} and Table \ref{tab3}).
The compact OH emissions would then be systemic emission regions that are re-amplified by excited gas within the dominant foreground structures at velocity 5314 \kms for Arp\,220E.

The curious aspect regarding the compact emission regions is that they appear as identifiable features only on the east and southeast sides of the main emission regions.
As discussed further in Section~\ref{sec6} below, this may be related to variation in velocity-coherent amplifying column density related to a velocity gradient in the foreground material.

The position-velocity diagram in Figure \ref{fig18} that the systemic components E1 - E3 show a gradient of 19 \kms over 107 pc (0.18 \kmspc).
Similarly, the gradient in the foreground screen is on the order of 18 \kms over a distance of 38 pc (0.47 \kmspc).
A difference exists in Arp\,220E where the 1667 MHz line width is almost uniform across the region, while the 1665 MHz extension emission shows a wide section towards the East followed by a narrower eastern edge (Figs. \ref{fig13} and \ref{fig14})

\begin{table*}
\begin{center}
\caption{OH line properties in MERLIN - EVN emission regions}
\begin{threeparttable}
\begin{tabular}{lccccccl}
\hline
Location\tnote{1}  &   Velocity & FWHM & S1667       & S1665       &  1667/1665 & Opt.Depth & Comment\\
                            &   (\kmss) & (\kmss)  & (mJy b$^{-1}$) &  (mJy b$^{-1}$)    & ratio         &$\tau_{67}$&\\
\hline
West North1       & 5258   & 29               & 2.9    & 0.3       &   9.7   &  -5.0   &  foreground \\
West North2       & 5268   & 34               & -1.5   & $<0.1$ & --        & --     & fg absorption  \\
                           & 5375   &  34             &  2.5   & $<0.1$ & $>$25  &$>$7.3  & W systemic\\
 W-Main North    &  5255  & 24               &  83   & $<$1.5 &  $>$55 &  $>$-9.0    & foreground \\
W-Main South    & 5245   & 165\tnote{2}& 27    & 6.0        &   4.5     & -3.0   & foreground; double peak \\
W1  -- North        & 5361   &  39              &  31   & $<$1.0 &  $>$31  &   $>$7.6   & W systemic\\
W1 -- North EVN & 5366  &  26              & 155   & (5)        & (31) & (7.6) & compact\tnote{3}\\
W2  -- SouthEast & 5317  &  68\tnote{2} &   7     &   0.8     &    8.7   &  -4.7  & W systemic\\
W2 -- SouthEast EVN & 5325 & 65        & 12      & --         & -- & -- &  compact\tnote{3} \\
W3  -- East         & 5366   & 39\tnote{2} &  12    &   4.0    &    3.0   &  -1.9  & W systemic\\
W3  -- East         & 5366   & 39\tnote{2} &  12    &   4.0     &    3.0   &  -1.9  & W systemic\\
W4 -- South        & 5246   & 19               & 4.2     &  0.6      &    7.0    &  -4.2 & foreground \\
                           &  5346  &   87             & 1.2    & 0.2        &     6.0   &   -3.8 & W systemic \\
W5 -- SouthWest & 5212 & 22               & 7.8     & 0.5/1.5 & 15.6      &   -6.1    & foreground\\
                           & 5296  & 32               & 5.0     & 1.5      & 3.3    & -2.1 & W systemic \\
WS1                   & 5255  & 24               &14.0   & 0.7    &  20.0     &  -6.8      &  foreground\\
WS2                   & 5255  & 124             & 4.0   & 1.1     &  3.6      &   -2.7     & foreground\\
                           & 5438  & 120             & 2.0   & 2.0     &  1.0     &  small & E systemic \\
  & & & & &&&\\
East -Main West & 5304 &  78  &   93   & 11.5    & 8.1     & -4.5    &  foreground \\
East -Main East & 5322  &   68  &  72    &  12.1   & 5.9     &  -3.7   &  foreground  \\
E1  -- East          & 5429  &   48 &   9      & 1.1      &  8.2    &  -4.5   & E systemic \\
E1  -- East  EVN & 5444  & 18 & 50 & -- & -- & -- & compact double\tnote{3}\\
                           & 5411   & 23 & 20 & -- &-- & -- & \\
E2  -- South        & 5410   &   39   & 13   & 1.0  &  13.0  &  -5.7   & E systemic \\
E2  -- South  EVN & 5410 & 35 & 42  & --  & -- & -- & compact\tnote{3}\\
E3  -- NorthEast    & 5429 &   53  &  4.5   & 0.3      & 15.0   &  -6.0  & E systemic  \\
 \hline
\end{tabular}
 \begin{tablenotes}
        \item[1]{Note 1: The locations of the emissions regions are identified in Figures \ref{fig15}, \ref{fig16}, and \ref{fig17}.}
         \item[2]{Note 2: Narrower profile superposed on a broad base ranging from 5000 to 5400 \kmss. }
         \item[3]{Note 3: These compact components have been detected in the EVN data as shown in Figures \ref{fig8} - \ref{fig10}. }
   	 \end{tablenotes}
\label{tab2}
  \end{threeparttable}
  \end{center}
    \end{table*}

\begin{figure*}
\begin{center}
\includegraphics[width=2.1\columnwidth,angle=0]{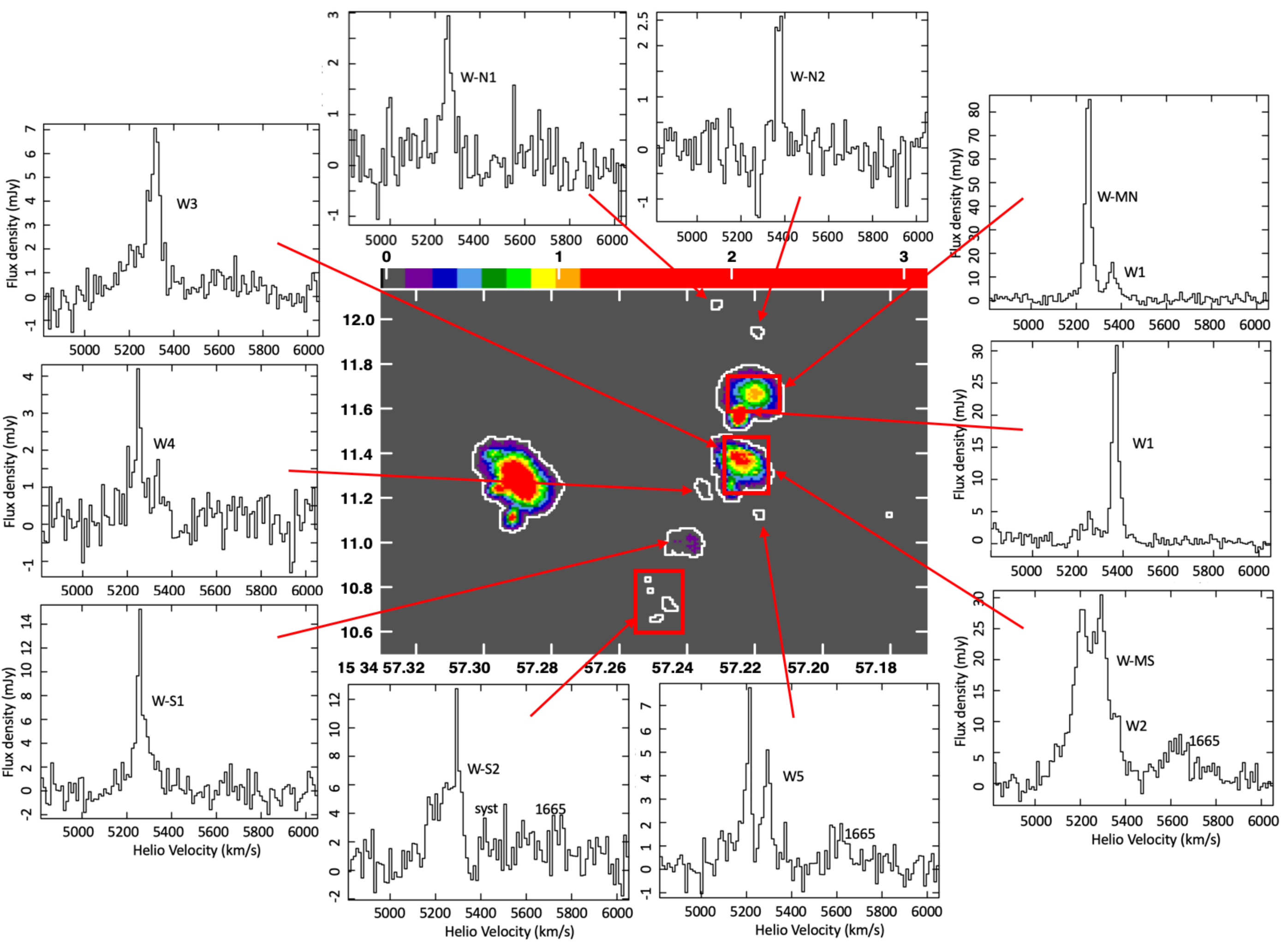}
\caption{Spectral components in Arp\,220W.
The velocity range of all spectra runs from 4809 to 6046 \kms increasing from left to right.
The vertical scale of all spectra is in mJy except that the scale is not equal in all spectra in order to show the weaker profiles.}
\label{fig17}
\end{center}
\end{figure*}

A comparison of the observed OH emission with the large-scale CO emission shows that the optical velocity of Arp\,220E is in agreement with the CO emission spectra \citep[Figure 1 in ][]{WheelerEA2020}. 
This spectrum shows two $^{12}$CO\,(3 - 2) emission components at optical-defined velocities of 5297 and 5603 \kms separated by a strong absorption at 5440 \kmss.
The systemic (optical) velocity of 5425 \kms for the OH emission coincides with the strong absorption feature at the central continuum source, while the velocity of the foreground material of 5314 \kms indicates an association with a low-velocity CO component. 
The high-velocity CO component at 5603 \kms has no OH counterpart and  appears to be located behind the nucleus.

The most recent 0Moment maps of the $^{12}$CO\,(3 - 2)  and the optically thin $^{13}$CO\,(4 - 3) emission data show an enhanced emission region at the location of Arp\,220E within a large scale emission structure drifting in northeast direction \citep{WheelerEA2020}. 
The $^{13}$CO\,(4 - 3) emission at the Eastern nucleus shows a SW-NE inclined and slightly warped disk structure extending to 366 pc and covering a radio velocity range from 5160 to 5680 \kms \citep{WheelerEA2020}. 
The systemic OH emission features at the East nucleus are consistent with these larger scale molecular structures, although they only highlight the central region of this disc that provides the FIR pumping emission.

For comparison, the formaldehyde maser emissions at the Eastern nucleus occur at the SW side of the nuclear centre and indeed show emission at velocities close to 5400 \kms  \citep{BaanEA2017}. 
However, the western H$_2$CO emission spectrum shows a profile that also encompasses the velocity range of the foreground component reaching down to below 5100 \kmss, which is in agreement with the foreground CO data. 
This would suggest that at Arp\,220E, the foreground component may also re-amplify the SF regions at the systemic velocity.

\subsection{Extended and compact OH emission at West}\label{sec43}

A deep 0Moment map of the 1667 MHz emission in Arp\,220 also shows the dominant emission region at the West nucleus, in addition to three compact region on its eastern side (Fig. \ref{fig15}).
Two of these regions have counterpart in the EVN data (Fig. \ref{fig8}) and are clearly visible in Figures \ref{fig13} and \ref{fig14} and are identified in Figure \ref{fig15} as W1 - W3. 
In addition to these, a number of additional emission regions may be identified to the north and south of Arp\,220W.

The spectral velocity components at all identified  emission regions are displayed in Figures \ref{fig16} and \ref{fig17}.
The compact components W1-W3 at the Western nucleus have a systemic velocity of about 5360 \kmss, which is 75 -140 \kms higher than the extended West OH components with an approximate velocity of 5254 \kmss. 
Similar to the situation at the East nucleus, the W1-W3 regions are also located on the east--southeast side of the extended emission regions (see Sect. \ref{sec5}).
In addition, the nearby components W4 and W5 both show a low-velocity component in agreement with that of the extended main regions, and a high-velocity component in agreement with those of the W1-W3 regions.  
Similarly, the W-S1 and W-S2 regions show emission at the foreground velocity, except that the W-S2 region shows an emission pair (with $R_H$ = 1667/1665 $\approx$ 1) at a velocity of 5438 \kmss, which corresponds to the systemic velocity of the East nucleus.
 The relative sizes, locations, and velocity offsets with the more extended OH emission regions suggest that any of these compact regions correspond to star formation regions belonging to the nuclear region of the underlying galaxy. 

\begin{table}
\begin{center}
\caption{Systemic Velocities in Arp\,220}
\begin{tabular}{lccccccl}
\hline
Arp\,220       &   Systemic & Foreground & Position   &  Comment\\
Nucleus 	   &   Velocity    &  Velocity      &  Angle     &            \\
                      &   (\kmss)     & (\kmss)      & (degree)    &   \\
\hline
West     & 5360            &     5354         & 168        &     nearly edge-on\\
 East     & 5425		   &	5314		  &  45	   &   edge-on    \\
 \hline
\end{tabular}
\label{tab3}
  \end{center}
    \end{table}

\begin{figure*}
\begin{center}
\includegraphics[width=1.4\columnwidth,angle=0]{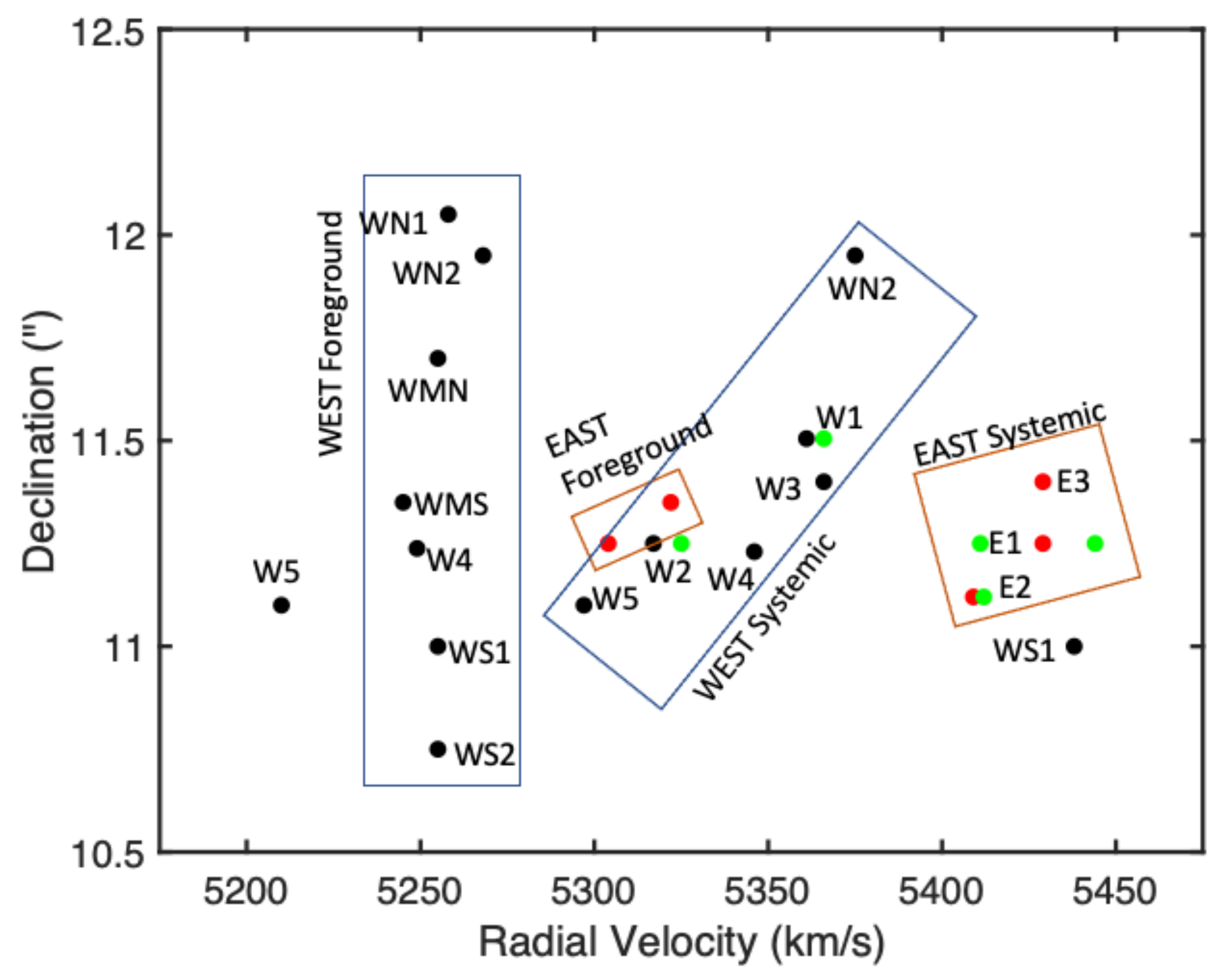}
\caption{Spectral components in Arp\,220.
The central velocities of the identified components of Arp\,220 are presented in a position-velocity diagram.
Black and red data points are found at the West and East nucleus, respectively, and green data points represent EVN detections at the two nuclei. 
The low-velocity data point of component W5 lies within the extended range of the broad foreground molecular CO structure.}
\label{fig18}
\end{center}
\end{figure*}

The spectra in the composite diagram of the West nucleus again shows the superposition of two structural components with distinct velocity systems  in Figures  \ref{fig17}. 
The position-velocity diagram in Figure \ref{fig18} depicts these velocity systems using the estimated velocity values obtained from the spectra at various locations (see Table \ref{tab2}).
The higher velocity compact components and surrounding emission components represent the systemic velocity of West, while the lower velocity components represent foreground gas structures.
The compact components at West identify a systemic velocity of 5360 \kms (see Fig. \ref{fig18}), while the 
dominant foreground structures would be at 5254 \kms (see Table \ref{tab3}.

The position-velocity diagram in Figure \ref{fig18} shows a clear velocity gradient for the systemic velocity components at the West nucleus.
Also taking into account the outlying components WN2 and W5, the small South to North gradient at the West nucleus has a velocity range of approximately 78 \kmss, and covering a distance of 324 pc (grad = 0.24 \kmspc). 
This gradient would confirm that the systemic components at the West nucleus are representative of rotation in a nearly edge-on disc.
A weak gradient is also seen for the foreground screen when considering the data points WS3 to WN1, which covers only 28 \kms over a distance of 800 pc (grad = 0.03 \kmspc). 
The foreground screen is drifting slowly in front of the nuclei in a NE direction.
The corresponding 1665 MHz emission in Arp\,220W-South shows an apparent gradient in the opposite direction.

A comparison with the existing large-scale CO observations shows that the optical velocities are in agreement with the emission spectrum at Arp\,220W \citep[Figure 1 of][]{WheelerEA2020}. 
This spectrum shows two $^{12}$CO\,(3 - 2) emission components at optical-defined velocities of 5245 and 5558 \kms separated by a strong absorption at 5424 \kmss.
The systemic velocity of the OH emission at the West nucleus 5360 \kms would coincide with this strong absorption feature associated with the central continuum source, while the velocity of the foreground material of  5254 \kms is associated with the low-velocity CO component. 
Again the high-velocity CO component at the nucleus Arp\,220W has no OH counterpart and appears to be located behind the nucleus.

The most recent $^{12}$CO\,(3 - 2) and optically thin $^{13}$CO\,(4 - 3) 0Moment maps show strongly enhanced emission at the central location of Arp\,220W within a large scale emission structure drifting at about 2.4 \kmspc in NE direction \citep{WheelerEA2020}. 
This emission region appears as a torus structure centred on the continuum emission and consistent with the large 2Moment velocity width seen at the West nucleus (Fig. \ref{fig13}). 
The 2Moment image of this elongated and tilted S-N torus structure suggests clockwise (East blue and West red) rotation with a width of about 480 \kms and an estimated orbital velocity of 240 \kmss.
The extended 1667 MHz OH emission regions have an almost South-North orientation and are separated by about 130 pc as they straddle the absorption gap centred on the peak of the radio continuum at Arp\,220W.
This suggests that these regions represent the tangential sections of the nuclear torus with an estimated outer diameter of 240 pc.
Similarly, the velocities of the OH SF components identified for Arp\,220W are consistent with the clockwise rotation of the inner CO structure and appear consistent with the N-S oriented SF regions being on the front side of a central torus.

The presence of a tilted N-S molecular torus in the nuclear region of Arp\,220W with a clockwise orbital motion and a central absorption component would also be consistent with the position-velocity behaviour of the HCN emission in the nuclear region without invoking an outflow and an E-W nuclear region \citep[see][]{BarcosMunozEA2018}. 

For comparison, the formaldehyde maser emissions at the West nucleus are thought to be associated with the disk component but the emission spectra also show multiple components at the velocity of the foreground  \citep{BaanEA2017}.
This suggests that the emission of the systemic regions is also re-amplified by the foreground structure.
A comparison of the spectrum of Centre and West H$_2$CO components at Arp\,220W shows that the Centre profile has a 130 \kms lower velocity and could also be associated with the front side of the torus.

\subsection{Outflows in Arp\,220}

Extended blue wings have been detected in single dish OH emission spectra in Arp\,220 and other OHMM that have been interpreted as outflows \citep{BaanHH1989}.
In Arp\,220 this blue wing may extend about 1000 \kms below the systemic velocity and some evidence for such an extension may even be found in the current EVN spectrum of W1 in Figure \ref{fig9}.
Similarly, recent CO observations of Arp\,220 show the broad (1300 \kmss) emission profiles at the two nuclei, where the outer blue and red parts of these profiles have also been designated as outflows
\citep{WheelerEA2020}.
However, considering that Arp\,220 is a strongly interacting system showing multiple CO emission regions along the line of sight, one may reconsider the nature of these outflows.
Are these really outflows from the nuclear regions or do they represent disk material flung away from the system at a larger relative velocity during this merger interaction ?
Maybe indeed the more likely explanation for these high velocity components is that they result from the interactive nature of the system.
The velocity in the CO data does not reveal the distance to the nuclear cores and with the right line-of-sight conditions any low-velocity foreground component can also amplify the radio background and produce a low-level blue wing in the OH profile.

\section{Two masering scenarios}\label{sec5}

The detailed interpretation of the Arp\,220 OHMM system shows a variety of emission properties and line ratios at the different regions within this interacting system, suggesting differences of the masering environments in the emission regions.
However, the basic scenario for the OH amplification is having an alignment of:
1) a radio background serving as seed radiation, 
2) an FIR pumping agent with the right spectral shape to create a population inversion, and
3) and an embedded or foreground column density with a line-of-sight velocity-coherent column density.
If all such conditions are fulfilled, the foreground molecular structures can amplify both an extended radio continuum background at its own radial velocity and re-amplify any maser emission originating in the underlying star-formation regions at their own radial velocity.

In a controlled environment under LTE circumstances, the optical depth of the 1667 and 1665 MHz OH transitions would vary as $\tau_{67}$ = $1.8 \times \tau_{65}$, which suggests that the line flux ratio varies as:
\begin{equation}
R_H = S_{1667} / S_{1665} = (e^{-\tau_{67}} - 1)/(e^{-\tau_{67}/1.8} - 1).
\label{eq1}
\end{equation}
While this ratio will be independent of the background (or seed) continuum radiation field behind the emission region, the variation of the amplifying optical depth across any masering region should always give a clear correlation between the two emission lines. 

The lower resolution MERLIN data show an $R_H \approx$  4.5 for both main East and West components suggesting relatively lower (integrated) optical depths (see Fig. \ref{fig3}). 
However, the 1667/1665 OH line ratios of components in the MERLIN-EVN data show a large range of $R_H$ values ranging from 3 to 20,  which indicates distinct differences in the masering conditions in the foreground gas and the systemic environments. 

The OH emission from the foreground gas varies with the varying amplifying gain across the face of the foreground structure convolved with that of the background radio structure.
As a result of these variable parameters, the amplifying optical depth of the foreground material is found to vary significantly with values between $\tau_{67}$= -2.7 and -6.8 with one apparent value $>$-9.0 in Arp\,220W-North.
Similarly, the emission of the systemic SF regions will result  from the intrinsic gain in the SF region and the gain provided by a foreground column with a similar velocity. 
The available data of SF-related features at the systemic velocity of the two nuclei shows an optical depth range of -1.9 to -6.0 with an extreme value of $>$-7.6 for the W1 region in Arp\,220W-North.
Based on these values, the foreground regions contribute a small addition to the gain for the systemic SF regions.

RADEX simulations \citep{vdTakEA2007} show that the range of optical depths in the OH main lines required in the foreground masering regions at each of the nuclei can be achieved when the molecular gas is relatively cold at $T_k$ = 20 K and is exposed to an FIR radiation field emitted by warm dust with $T_{dust}$ = 50 K. 
The required OH column densities are on the order of 10$^{17}$ cm$^{-3}$.
The masering conditions provide some diagnostics of the foreground material.

The systemic star-formation regions are curiously located at the east-southeast edges of the main emission regions at each of the two nuclear regions. 
The reason for this may relate to a velocity gradient in the $^{12}$CO(1-0) foreground appears to run globally from southwest to northeast  \citep{WheelerEA2020,SakamotoEA2008}.
Since foreground re-amplification requires an sufficient inverted column density at the exact velocity of the systemic SF regions, the foreground velocity gradient and the density distribution in the foreground appear to provide a certain optical depth for the regions at the eastern edge of the main emissions and not for the distinct SF regions at the western side of the nuclei. 

\section{Discussion}\label{sec6}

The OH MegaMaser activity in Arp\,220 appears to be more complex than was anticipated on the basis of earlier observations.
While the lower resolution MERLIN data showed the larger scale foreground emission regions and did not distinguish emission components at the galactic cores, the high-resolution EVN data resolved the extended emission and detected a few of the high-brightness star-formation components at each of the galactic cores.
However, the combined MERLIN - EVN data with intermediate resolution provides a much clearer view of the structural components in Arp\,220.
In particular, the combined data provides a consistent masering scenario, where the FIR-pumped foreground material amplifies the background continuum from within the galaxy core regions and independently re-amplifies the galactic SF-related components.

For the case of the OH emission in Arp\,220 a clear velocity distinction can be made between the compact star-formation regions at the systemic velocity of the East and West nuclei and a foreground screen covering the nuclear regions at velocities about 100 \kms below that of the nuclei.
Other higher-velocity molecular CO components do not have a counterpart in the OH data and appear to be located behind the nuclei \citep[see][]{WheelerEA2020}.
Assuming that the velocity of the various compact emission regions in the EVN and combined MERLIN-EVN data are part of the underlying galactic nuclei, these components accurately determine the systemic velocities of the nuclei of 5425 \kms for the East nucleus and 5360 \kms for the West.
As expected, these systematic velocities of the two nuclei correspond closely with those of the apparent absorption regions in the large-scale molecular structures \citep{WheelerEA2020}.
Subsequently, the velocities of the amplifying foreground regions that produce the bulk of the OH emission are at about 5312 \kms for the East region and at 5260 \kms for the West region, which is 100 \kms below the systemic velocities of the two nuclei.

Because the OH emission from the foreground dominates, very little of the systemic structures of Arp\,220 can be detected except for the evidence that large scale star-formation related FIR emission serves as a pumping agent for the foreground material.
The East nucleus of Arp\,220 appears to have a SW-NE orientation at a position angle of about 45$^{\circ}$ and shows a small velocity gradient in that direction. 
The systemic components at the West nucleus suggest an edge-on S-N orientation with a position angle of -12$^{\circ}$, which is consistent with the apparent presence of a nuclear torus structure seen at the nucleus within the large scale $^{12}$CO\,(3 - 2) emission data. 
No evidence can be found in the OH data for an E-W orientation of the West nuclear region.
This nearly edge-on torus appears to have a clockwise rotation with an estimated orbital velocity of about 100 \kmss, which would be consistent with the presence of compact SF-related emission regions on the eastern edge of the OH emission region.

The maser amplification scenario proposed early for the OH MegaMaser emission is found to be clearly applicable for Arp\,220 \citep{Baan1985,Baan1989}.
This scenario represents a line-of-sight convolution of the variable amplifying optical depth in the foreground gas with the distributed source of molecular pumping and the radio emission in the background.
Prominent MegaMaser emission lines of other molecules in extragalactic sources are likely to be generated in a similar manner. 
The H$_2$COMM emission components in Arp\,220 are similarly superposed on the core regions of the galaxies and show the velocity range of both the systemic and the foreground regions \citep{BaanEA2017}.
Similarly, for prominent H$_2$OMM sources, such as NGC\,4258 and NGC\,1068, the collisionally excited gas is also superposed on the continuum structures \citep{HerrnsteinEA1999,GallimoreEA2004,BaanEA2022}.
However, this type of amplified emission always depends on a geometry where the amplifying column density is aligned with a background continuum.
For extragalactic sources, the probability of this happening may be low and may vary significantly for different molecules.
For Galactic sources, this geometry requirement may account for the (non-)occurrence of maser action in certain environments but it may also explain the variability and dynamic behaviour observed in sources.

The dominance of the OH emission from the foreground material highlights the complex nature of the nuclear regions of Arp\,220 with large amounts of molecular material at velocities below and above the systemic velocity of the system. 
The large line-of-sight velocity width of the CO emission in Arp\,220 of some 800 \kms \citep{WheelerEA2020} appears to be a characteristic for violent galaxy mergers as has been found in higher redshift OHMMs with OH line widths as high as 2400 \kms \citep{BaanEA1992,DarlingG2002,PihlstromEA2005}.
Incidentally, the presence of blueshifted foreground gas may also explain the blue tails in the OH emission profiles (interpreted as outflows) reaching some 800 \kms as in the case of Arp\,220 \citep{BaanHH1989}.

\section{acknowledgement}

WAB acknowledges the support from the National Natural Science Foundation of China under grant No.11433008 and the Chinese Academy of Sciences Presidents International Fellowship Initiative under grants No. 2021VMA0008 and 2022VMA0019.
TA acknowledges the support of the National SKA Program of China (grant  2022SKA0120102). 

The European VLBI Network is a joint facility of independent European, African, Asian, and North American radio astronomy institutes.
MERLIN is a National Facility operated by the University of Manchester at Jodrell Bank Observatory on behalf of STFC.

\section{Data availability}
The data for the experiments of Arp\,220 may be obtained from the MERLIN Data Archive under project code MN-03B-22 and the EVN Data Archive under project code EB022C.
Calibration of the data has been done using the NRAO Astronomical Image Processing System (AIPS) and with the ATNF MIRIAD Data Reduction Software for self-calibration and imaging.




\end{document}